%% file: main.tex
\documentclass[%
reprint,
superscriptaddress,
groupedaddress,
amsmath,amssymb,
aps,
pra,
]{revtex4-2}

\usepackage{graphicx}
\usepackage{dcolumn}
\usepackage{bm}
\usepackage{xcolor}
\usepackage{qcircuit}
\usepackage{braket}
\usepackage{dsfont}
\usepackage{placeins}
\usepackage{comment}
\usepackage[utf8]{inputenc}
\usepackage{pgfplots}
\usepackage{hyperref}
\usepackage{cleveref}
\usepackage{cleveref}
\usepackage{array}
\usepackage{tabularx}
\usepackage{booktabs}

\DeclareUnicodeCharacter{2212}{−}
\usepgfplotslibrary{groupplots,dateplot}
\usetikzlibrary{patterns,shapes.arrows}
\pgfplotsset{compat=newest}

\begin{document}

\title{Scalable Fluxonium-Transmon Architecture for Error Corrected Quantum Processors}

\author{Lukas Heunisch\textsuperscript{1}}
\email{lukas.heunisch@fau.de}
\author{Longxiang Huang\textsuperscript{2,3}}
\author{Timo Eckstein\textsuperscript{1}}
\author{Stephan Tasler\textsuperscript{1}}
\author{Johannes Schirk\textsuperscript{2,3}}
\author{Florian Wallner\textsuperscript{2,3}}
\author{Verena Feulner\textsuperscript{1}}
\author{Bijita Sarma\textsuperscript{1}}
\author{Klaus Liegener\textsuperscript{2,3}}
\author{Christian M. F. Schneider\textsuperscript{2,3}}
\author{Stefan Filipp\textsuperscript{2,3,4}}
\author{Michael J. Hartmann\textsuperscript{1}}
\email{michael.j.hartmann@fau.de}

\affiliation{\textsuperscript{1}Physics Department, Friedrich-Alexander-Universität Erlangen Nürnberg, Erlangen, Germany}
\affiliation{\textsuperscript{2}Walther-Meißner-Institut, Bayerische Akademie der Wissenschaften, 85748 Garching, Germany}
\affiliation{\textsuperscript{3}Technical University of Munich, TUM School of Natural Sciences, Department of Physics, 85748 Garching, Germany}
\affiliation{\textsuperscript{4}Munich Center for Quantum Science and Technology (MCQST), Schellingstraße 4, 80799 München, Germany}

\date{\today}

\begin{abstract}
We propose a hybrid quantum computing architecture composed of alternating fluxonium and transmon qubits, that are coupled via transmon tunable couplers. We show that this system offers excellent scaling properties, characterized by engineered zero $ZZ$-crosstalk in the idle regime, a substantial reduction of level-crowding challenges through the alternating arrangement of different qubit types within the lattice, and parameter regimes that circumvent the capacitive loading problem commonly associated with fluxoniums. In numerical simulations, we show a parametrically driven CZ-gate that achieves a closed-system infidelity that is orders of magnitude below the coherence limit for gate durations $\gtrsim 30\,\rm{ns}$ using a two-tone flux pulse on the tunable coupler. Furthermore, we show that this gate scheme retains its fidelity in the presence of multiple spectator qubits, making it a scalable solution for large lattices. Moreover, for the implementation of error correcting codes, our approach can leverage the long coherence times  and large non-linearities of fluxoniums as data qubits, while fixed-frequency transmons with established readout techniques can serve as measurement ancillas.
\end{abstract}

\maketitle

\input{./1_Introduction.tex}
\input{./2_Circuit.tex}
\input{./3_CZ_gate.tex}
\input{./4_Larger_system.tex}
\input{./5_Conclusions.tex}

\appendix
\input{./Appendix.tex}

\FloatBarrier

\nocite{*}
\bibliography{mybib}

\end{document}

%% file: 1_Introduction.tex
\section{\label{sec:intro}Introduction}
Superconducting qubits have emerged as a highly suitable platform for realizing quantum computing, where architectures involving a significant number of qubits are mostly based on Transmon designs \cite{AA, AB}. Transmons are highly valued for their simplicity, which makes them easy to fabricate, to control and to scale. With increasing qubit number, Transmon based architectures however appear to face challenges. Their comparatively strong coupling to external fields limits their coherence times to typically tens or hundreds of microseconds \cite{AA, AB} and their small anharmonicity $\sim -250\,\rm{MHz}$ limits the speed of gate operations since strong, spectrally broad pulses can cause population of leakage states.  In addition, they face a risk to couple to two-level-defects \cite{AC}, which can change transition frequency from one cool-down to another.

\begin{figure}[!htp]
    \centering
    \includegraphics[width=\linewidth]{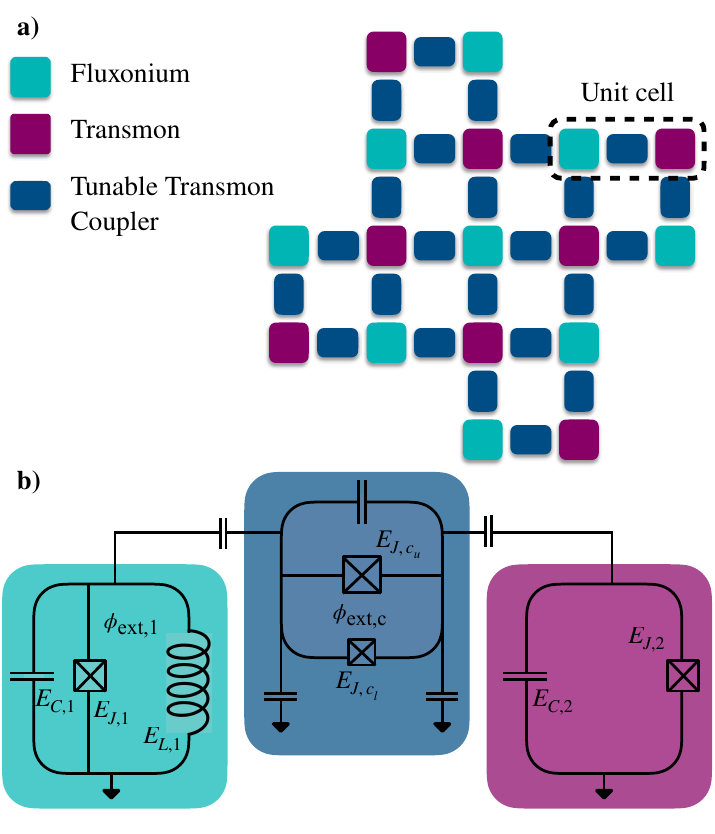}
    \caption{a) Schematic of a distance-three surface code leveraging a hybrid qubit architecture. The cyan squares represent Fluxoniums, which can be used as data qubits, whereas the purple squares are Transmons, which thus can be used as measurement qubits. b) Lumped element circuit of a unit cell in the proposed system comprising one Fluxonium (left), one Transmon (right) and one tunable coupling circuit (middle) connecting the two qubits.}
    \label{fig:circuit}
\end{figure}

Fluxoniums \cite{BA, BF} have favorable properties to overcome some of these challenges and are therefore considered a promising candidate for a next generation of qubits. On top of long coherence times, on par or even exceeding those of Transmons \cite{BL} and a reduced sensitivity to two-level systems, due to their small charge matrix elements \cite{BF, BP}, their large anharmonicity in the GHz-regime suppresses the population of leakage states when the qubit is strongly driven. Recent research has focused on exploring interactions via capacitive \cite{BB, BC, BD, BE, BG, BN, BP, BR, BS, BV, BW, BH} or inductive \cite{BJ, BK, BM, BQ, BS, BU} coupling using direct coupling, tunable couplers or resonators. While these studies have shown significant promise, current experiments are limited to coupling just two Fluxoniums. 

Scaling Fluxonium architectures remains a significant challenge, as the substantially larger capacitive energy of Fluxonium qubits, compared to Transmons, restricts the size of coupling capacitances that can be employed to connect to four neighboring qubits in a square lattice architecture. This restriction, commonly referred to as capacitive loading, imposes practical limits on many gate schemes that are currently under investigation \cite{BE}. In addition, the denser energy spectrum of Fluxonium qubits along with the absence of strict selection rules increases frequency crowding, which in turn enhances the risk of crosstalk during single-qubit gates and can lead to phase errors arising from strong interactions between higher excited states \cite{BI, BV}.

Moreover quantum computations require sufficiently performant quantum error correction. The surface code \cite{DM, DH, DA} is considered the most promising candidate for superconducting circuits since it is well suited for two-dimensional lattice geometries and has a high error threshold. Multiple studies have focused on incorporating it into superconducting circuits \cite{AB, DB, DC, DD, DE, DF, DG, DI, DJ, DK, DL, CC, DO} with the aim of improving gate fidelities to the level required for fault tolerance. However, the surface code requires a significant qubit overhead, posing immense scaling challenges for superconducting quantum hardware \cite{CA, CC}.

In this work, we address these challenges by proposing a novel hybrid architecture that combines Fluxonium and Transmon qubits and is highly suited for a surface code implementation, see \autoref{fig:circuit} for an example of a distance-three surface code architecture. In our scheme, high-coherence Fluxoniums are used as data qubits, while Transmons serve as ancillas or measurement qubits. While the opposite choice is also possible, the proposed allocation is advantageous because the Fluxonium’s large anharmonicity provides strong protection against leakage, which is particularly important for data qubits, as they cannot be reset during the execution of a quantum circuit. 

We furthermore demonstrate that our proposed architecture exhibits excellent scaling properties. Firstly, by operating in a suitable, experimentally accessible parameter regime, we circumvent capacitive loading, enabling scalable square lattice connectivity. Furthermore, we show that $ZZ$-crosstalk can be fully suppressed by using a tunable Transmon coupler that capacitively couples to the Fluxoniums and Transmons. Finally, the alternating arrangement of different qubit types significantly mitigates frequency crowding. Since Fluxoniums and Transmons operate on distinct energy scales, this architectural feature naturally reduces the risk of single-qubit gate crosstalk and contributes to overall system robustness. Our simulations also show the possibility of implementing fast and high-fidelity parametrically driven CZ-gates. We find that the degradation of the gate fidelity due to spectator qubits saturates already for as little as two spectator qubits and well above the decoherence limit, highlighting the potential of this architecture.

%% file: 2_Circuit.tex
\section{\label{sec:circuit}Circuit}
\subsection{Lumped element Hamiltonian}
The circuit we examine comprises a Fluxonium qubit and a Transmon qubit, which are capacitively coupled via an asymmetric tunable Transmon coupler (see \autoref{fig:circuit} b). The Transmon is chosen to be a fixed-frequency qubit to reduce its susceptibility to flux noise, thereby enhancing its coherence time. The circuit is described by the Hamiltonian 
\begin{align}
\label{eqn:Hamiltonian}
    H &= \sum_{i,j = 1, c, 2} 4E_{C, ij} \, n_i n_j \\ 
    &+ \frac{1}{2} E_{L,1} (\phi_1 - \phi_{\rm{ext},1})^2 - E_{J,1}\cos \left( \phi_1 \right) \notag \\
    &- E_{J,c_u} \cos \left( \phi_c \right) -E_{J,c_l} \cos \left( - \phi_c - \phi_{\mathrm{ext},c} \right) \notag \\
    &- E_{J,2} \cos \left( \phi_2 \right) \notag
\end{align}
where $\phi = 2\pi \Phi/\Phi_0$ corresponds to the reduced flux variable and $n$ is its canonical conjugate Cooper-pair number operator, see Appendix \ref{sec:circuitderivation} for further details. $E_J$, $E_C$, and $E_L$ denote the Josephson, charging, and inductive energies, respectively, as depicted in \autoref{fig:circuit}. A precise choice of the irrotational gauge to accommodate the time-dependent external flux would lead to only negligible corrections here \cite{EA}.

Using the standard quantization procedure, we can write the Transmon and the coupler as duffing oscillators with bosonic creation and annihilation operators with commutation relations $[b_i, b_j^\dagger] = \delta_{ij}$. Since the harmonic oscillator basis does not align close to the Fluxonium eigenstates, we express the Hamiltonian of the Fluxonium in its eigenbasis states. The total Hamiltonian is therefore given by
\begin{align}
    \label{eqn:FluxoniumH}
    H_F &= \sum_i \omega_F^i \ket{i}\bra{i} \\
    H_C &= \omega_c(\phi_{\text{ext},c}) a_c^\dagger a_c + \frac{\alpha_c}{2}a_c^\dagger a_c^\dagger a_c a_c \\ 
    H_T &= \omega_2 a_2^\dagger a_2 + \frac{\alpha_2}{2}a_2^\dagger a_2^\dagger a_2 a_2 \\
    \label{eqn:quantH}
    H &= H_F + H_C + H_T + \sum_{i < j} 4E_{C,ij} n_i n_j
\end{align}
In the following simulations, we consider this Hamiltonian, truncating each node, i.e. Fluxonium, coupler, and Transmon, to five energy levels, where the convergence of this truncation has been tested. The dressed eigenstates of the Hamiltonian in \autoref{eqn:quantH} contain the experimentally relevant computational basis and are denoted as $\ket{ijk}$, where the indices $i$, $j$, and $k$ correspond to the eigenstates of an isolated Fluxonium, coupler, and Transmon, for which the overlap of the direct product of these eigenstates (i.e. $\ket{ijk}_0$) with a state $\ket{ijk}$ is maximal, see also \cite{EC}.

\subsection{Parameter regime}
The challenge in identifying a suitable parameter regime for the Fluxonium qubit lies in balancing long coherence times with favorable scaling properties. We find that a significant anharmonicity of the Fluxonium is beneficial for the fast gate scheme that we propose in \Cref{sec:CZ_gate}, since it ensures that its second excited state is close to the Transmon transition frequency. We furthermore aim to operate the Fluxonium in a regime where single-qubit gates can still be performed via microwave control. This means, we avoid excessively low transition frequencies in the order of tens of MHz or even lower, where flux noise would become dominant and lead to a reduction in $T_1$ times \cite{FI}. In this low frequency regime, the Fluxonium becomes strongly noise-biased \cite{EG,EH}. 

We chose a parameter regime with a reasonably high Fluxonium frequency of $300\,\rm{MHz}$ and an anharmonicity of $3.7\,\rm{GHz}$. This choice requires the capacitive energy $E_C$ and inductive energy $E_L$ to be sufficiently large. Yet, a large $E_C$ requires designing a small shunt capacitance, which limits the ability to use large coupling capacitances without encountering problems with capacitance budgeting. A large inductive  energy, in turn, requires a very small inductance, which is prone to flux noise that can significantly reduce coherence time. Our choice balances these needs. The full set of circuit parameters used in the simulations result in the properties denoted in \autoref{tab:specs}. More information is provided in Appendix \ref{sec:parameters}. 
\begin{table}[!htp]
    \centering
    \begin{tabular}{c|c c c}
        \toprule
         & Fluxonium & Coupler & Transmon \\ \midrule 
        $\omega/2\pi$ & $300\,$MHz & $3.842\,$GHz & $4.4\,$GHz \\
        $\alpha/2\pi$ & $3.7\,$GHz & $-428\,$MHz & $-194\,$MHz \\
        $T_1$ & $825\,\mu$s & $18.5\,\mu$s & $70\,\mu$s\\
        $T_2$ & $149\,\mu$s & $33.5\,\mu$s & $103\,\mu$s\\
        $g_{101 \rightarrow 110}/2\pi$ & & \multicolumn{2}{c}{$-56\,$MHz}\\
        $g_{110 \rightarrow 200}/2\pi$ & \multicolumn{2}{c}{$58\,$MHz} & \\
        \bottomrule
    \end{tabular}
    \caption{Parameters of the architecture in \autoref{fig:circuit} b). The values for the Fluxonium were calculated from the energies in \autoref{eqn:FluxoniumH} using $\omega = \omega_F^1 - \omega_F^0$ and $\alpha = \omega_F^2 - 2\omega_F^1+\omega_F^0$. The estimates for the $T_1$ and $T_2$ times are based on \cite{BB, FA, FB}.}
    \label{tab:specs}
\end{table}

For the neighboring Transmon qubit, we set its frequency to $4.4\,\rm{GHz}$ such that it aligns closely with the second excited state of the Fluxonium. A relatively low $E_C$ value of $194\,\rm{MHz}$ ensures that it remains well within the Transmon regime ($E_J/E_C \gtrsim 70$).
Finally, the coupler transition frequency is set to lie between those of the Fluxonium and the Transmon. This configuration allows for parameter sets where $ZZ$-crosstalk $\zeta$, defined as
\begin{equation}
\label{eqn:zeta}
    \zeta = E_{101} - E_{100} - E_{001} + E_{000} \ ,
\end{equation}
is entirely eliminated and stays below $1\,\rm{kHz}$ over a wide range of coupler frequencies, varying over more than $100\,$MHz. Here, $E_{ijk}$ corresponds to the energy of the computational basis states $\ket{ijk}$. 

The suppression of $ZZ$-crosstalk can be understood by evaluating $\zeta$ defined in \autoref{eqn:zeta} perturbatively up to fourth order. Because the couplings between the qubits and the coupler are much stronger than the direct qubit–qubit coupling, the leading contributions to $\zeta$ arise from second-, third-, and fourth-order terms. In our floating-coupler architecture, however, the Transmon couples to the coupler with an opposite sign compared to the Fluxonium–coupler coupling. By adjusting the coupler frequency accordingly, the dominant $ZZ$ terms in this expansion therefore cancel out (see Appendix \ref{sec:ZZcrosstalk} for further details).

To maximize coherence, asymmetric junctions are used in the SQUID loop of the coupler, which allows to initialize the coupler at its lower sweet spot. The parameters are selected such that this lower sweet spot coincides with the point of complete $ZZ$-crosstalk suppression. While this approach is susceptible to fabrication errors, our simulations show that $ZZ$-crosstalk can still be entirely suppressed by readjusting the external magnetic flux over a wide range of fabrication imprecision, see Appendix \ref{sec:robustness} for further details.

The coupling capacitances in our simulation are also asymmetric. The capacitive coupling of the Fluxonium to the tunable Transmon coupler is small enough to avoid the problem of capacitance budgeting when scaling the system. Yet, the large coupling matrix element $\bra{1}n\ket{2}$ of the Fluxonium can compensate for the small capacitance value in gate operations, resulting in a coupling strength that is nearly as large as the coupling strength between the Transmons. The coupling capacitance between the Transmon qubit and the coupler is in our approach maximized while ensuring the delocalization $\epsilon_{i0k}$ of all computational basis states 
\begin{align}
    \epsilon_{i0k} &= 1 - \left| \braket{i0k | i0k}_0 \right|^2 \quad \text{for} \quad i,k \in \{0,1\}
\end{align}
remains below $1\,\%$. Here, the state $\ket{ijk}_0$ refers to the closest bare state, meaning the state that is a direct product of eigenstates of isolated or uncoupled circuit elements (Fluxonium, coupler and Transmon).

%% file: 3_CZ_gate.tex
\section{\label{sec:SQ-gates} Single-Qubit Gates}
We first show that it is possible to implement two distinct types of single-qubit gates for Transmons and Fluxoniums, and that these gates can run in parallel with very low crosstalk. In the first case, we use a microwave drive line, that we couple via a coupling capacitance, to inject pulses into the qubit. When driven near the qubit frequency they induce Rabi oscillations between $|0\rangle$ and $|1\rangle$. The resulting drive Hamiltonian $H_{D_T}$ is given by 
\begin{align}
    H_{D_T} =& -i\Omega g(t) \cos (\omega_d t) (a + a^\dagger ) \\
    \label{eqn:flattop_gaussian}
    g(t) =& \frac{1}{4}\left( 1 + \text{erf} \Bigl( \frac{t+\tau}{\tau}\Bigr) \right) \notag \\ 
    &\quad \left( 1 + \text{erf} \Bigl( \frac{t_{\text{gate}} - t - \tau}{\tau}\Bigr) \right) \ ,
\end{align}
where the parameters $\Omega$, $\omega_d$, and $\tau$ are optimized. We measure the precision of the gates via the process fidelity, that is defined by  \cite{FH, FJ} 
\begin{align}
    \label{eqn:infidelity}
    f_X = \frac{ |\text{Tr} (U^\dagger X)|^2 + |\text{Tr} (U^\dagger U)|}{\text{dim}(U)\cdot (\text{dim}(U)+1)}
\end{align}
We simulate a $10\,\rm{ns}$ $X$-gate and obtain a closed-system infidelity of $\epsilon_X = 1 - f_X = 6.5\times 10^{-5}$. 

For Fluxoniums, the flux line can be used to implement a high-fidelity $X$ gate. This reduces the overall circuit complexity and thereby decreases the number of potential noise sources. The corresponding drive Hamiltonian $H_{D_F}$ is given by 
\begin{align}
    H_{D_T} = g(t)\cos(\omega_d t) E_{L,1} \phi_1^{\text{ZPF}} (a + a^\dagger )
\end{align}
Using an optimization analogous to the one above, we simulate a $10\,\rm{ns}$ $X$-gate and obtain a closed-system infidelity of $4.2\times 10^{-7}$.  

To demonstrate that the alternating level structure in the proposed Fluxonium–Transmon lattice suppresses single-qubit-gate crosstalk, we simulate two simultaneous single-qubit gates embedded in the proposed Fluxonium--Transmon--Transmon (FTT) architecture. The resulting error of the effective two-qubit unitary is $1.3\times 10^{-4}$, which is a bit higher compared to the Transmon single qubit $X$-gate due to a slightly increased rest population of the state $\ket{002}$. However, simulating an $X$-gate on the Transmon with a Fluxonium spectator does not reveal any significant rest-populations of Fluxonium states, showing the strong decoupling of both qubits.

\section{\label{sec:CZ_gate}Parametrically driven CZ-Gate}
\begin{figure*}[!htp]
    \centering
    \includegraphics[width=\textwidth]{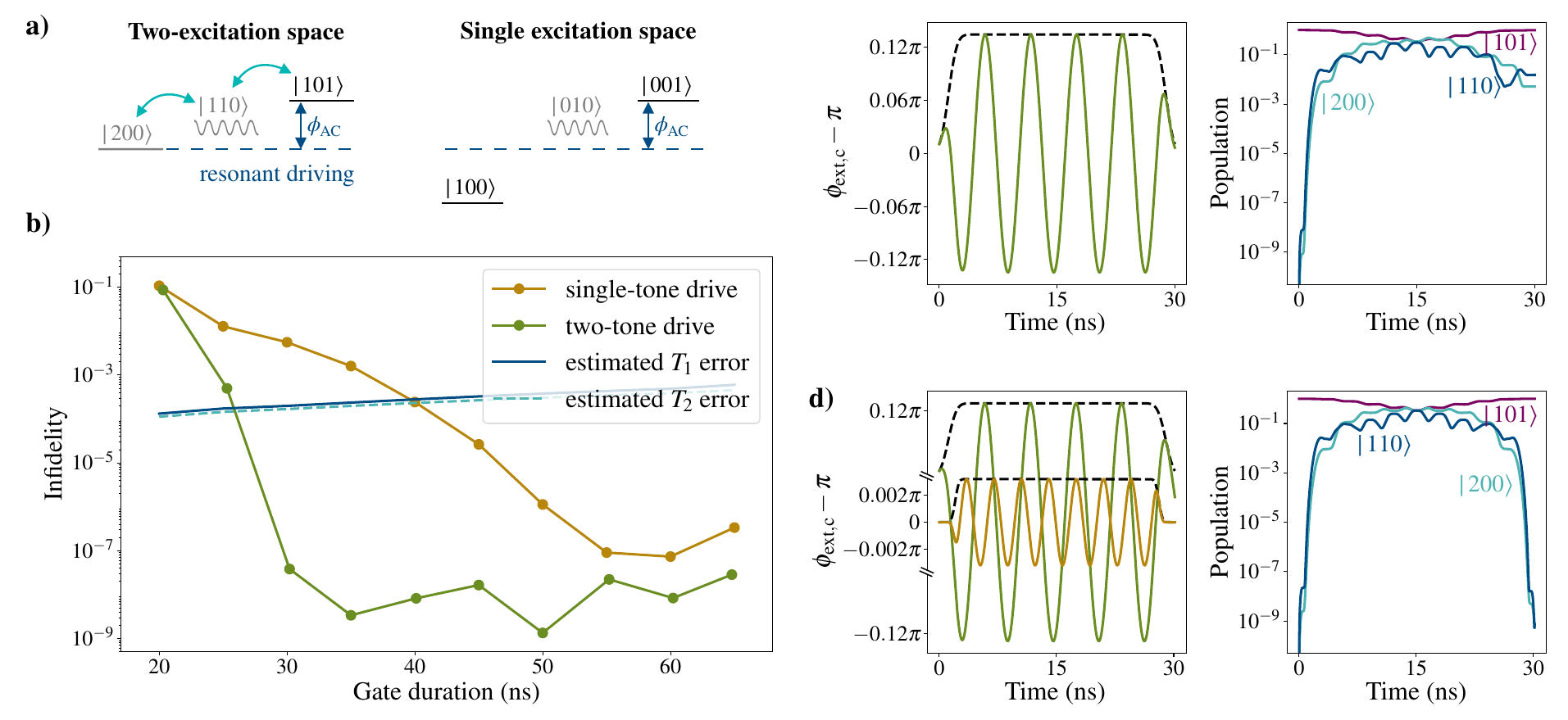}
    \caption{a) Gate scheme: The tunable coupler is driven with an ac flux pulse to drive Rabi oscillations between $\ket{101}$ and $\ket{200}$. In the two-excitation subspace (left panel), this modulation bridges the energy gap between the computational state $\ket{101}$ (black) and the non-computational state $\ket{200}$ (gray), where the excitations move via the coupler (state $\ket{110}$). In the single excitation subspace, transitions are off resonance (right panel). b) Gate infidelities for different gate times. The light brown curve shows results for single-tone control pulses, whereas the green line show results for two-tone control pulses. The dark blue line indicates the estimated error due to decoherence processes. c) - d) Flux-pulses for the simulation of a 30\,ns gate on the left and the state population of the input state $\ket{101}$ on the right. For the simulation in c) a single-tone flux pulse was used while the plots in d) result from a two-tone flux-pulse. The second tone of the two-tone pulse is much smaller as indicated by the broken vertical axis in the lower left panel, but can be used to suppress the leading leakage state contributions by more than four orders of magnitude.}
    \label{fig:infidelities}
\end{figure*}

To implement a fast and high-fidelity parametrically driven CZ-gate, as studied in e.g. \cite{FD, FE, FF, FG} we parametrically modulate the SQUID loop of the tunable coupler with an ac flux pulse. As depicted in \autoref{fig:infidelities}a, a resonant flux-pulse modulation of the coupler induces Rabi oscillations between the $\ket{101}$ state with an excitation in both, the Fluxonium and the Transmon, and the $\ket{200}$ state, where the Fluxonium is in its second excited state. This resonance occurs when the coupler’s modulation frequency is chosen close the energy difference $|E_{101}-E_{200}|$. Since the coupler is positioned at the lower sweet spot, a cosine-shaped flux pulse requires approximately just half of this frequency. This consequence of applying the drive directly to the non-linear element enhances the resilience of the devices against flux noise, since a low-pass filter can then protect the qubit from noise at the frequency of the qubit and decoherence caused by the control line \cite{BT}. 

Furthermore, in the proposed gate scheme the Fluxonium remains at its lower sweet spot, which eliminates the need for flux-tunability in the Transmon qubit and increases its coherence time compared to a tunable Transmons \cite{BJ, FC}. This offers two benefits: spectator qubits in a larger chip architecture remain decoupled (see \Cref{sec:spectator} for a detailed analysis), which is highly beneficial for scaling purposes \cite{AB, EC}, and errors due to decoherence are minimized.

In the following, we first discuss a simplified and perturbative approach for providing a motivation and intuitive picture for the gate. This is then complemented by a numerical analysis based on the full lumped element model in \autoref{eqn:Hamiltonian}, which provides a more precise quantitative assessment.

\subsection{Analytical Motivation}
Our goal is to drive a resonant transition between the $|101\rangle$ and $|200\rangle$ states by applying an ac flux pulse to the coupler. Including the mediating state $|110\rangle$, which is also the dominant leakage state in our scheme, we simplify the complete model of \autoref{eqn:Hamiltonian} to a three-level system. To describe the gate mechanism, we thus examine the drive Hamiltonian 

\begin{equation}
\label{eqn:H_D}
    H_D = \left( E_{J_\Sigma} \sqrt{\cos^2 \frac{\phi_{\text{ext,c}}}{2} + d^2\sin^2 \frac{\phi_{\text{ext,c}}}{2}} - E_{J_\Delta}\right) \cos \phi_c \ ,
\end{equation}
where $d = E_{J_\Sigma}/E_{J_\Delta}$ is the ratio of the sum $E_{J_\Sigma}$ and difference $E_{J_\Delta}$ of the Josephson junction energies in the SQUID loop of the coupler. We choose the external flux to consist of a dc and an ac component, leading to the phase, 
\begin{equation}
\label{eqn:phiext}
    \phi_{\text{ext,c}} = \pi + \phi_{AC} \cos (\omega_Dt) \ .
\end{equation}
Applying the Jacobi-Anger expansion to the square root and approximating it with a Taylor series, see Appendix \ref{sec:calculations}, we express the bracket of \autoref{eqn:H_D} as 
$ \alpha + \beta \cos (2\omega_D t)$, where the explicit forms of the constants $\alpha$ and $\beta$ are derived in Appendix \ref{sec:calculations}. Truncated to the three level system formed by the states $\ket{101}, \ket{110}$ and $\ket{200}$, our model thus reads 
\begin{equation}
\label{eqn:3lvlH}
    H_{3\text{\,level}} = \begin{pmatrix}
        E_{101} & & \\
        & E_{110} & \\
        & & E_{200}
    \end{pmatrix} + \mathbf{A} \cdot \bigl(\alpha + \beta \cos (2\omega_D t)\bigr)
\end{equation}
where all off-diagonal elements of the first matrix are zero and the elements of matrix \textbf{A} are given by $A_{ij} = \bra{i} \cos \phi_c \ket{j}$. To obtain an intuitive understanding of the gate, these can be derived via a static perturbation theory, which is explicitly done in Appendix \ref{sec:calculations}. For a higher precision of our simulation results, we however determine the elements of \textbf{A} as well as the constants $\alpha$ and $\beta$ numerically and use these values in the following.

Following the approach of \cite{ED}, we describe the system’s time evolution using a kick operator $\mathcal{K}(t)$, which kicks the system into a basis, where the long-term dynamics is generated by a time-independent effective Hamiltonian $H_{\text{eff}}$. Both $H_{\text{eff}}$ and $\mathcal{K}(t)$ can be derived perturbatively, assuming the energy splitting of the Hamiltonian is small compared to the drive frequency $\omega_D$. To describe the scenario of resonant driving, we apply a time-dependent rotation $R(t)$ as suggested in \cite{EE}, which eliminates energy splittings on the scale of $\omega_D$, see Appendix \ref{sec:calculations} for further details.

Fixing the drive amplitude for the system under study, we use this approach to calculate the corresponding drive frequency $\omega_D$ and gate time $t_{\text{gate}}$. This method already describes the gate with rather good accuracy for weak drive strengths. Compared to the value for the drive frequency $\omega_{\text{opt}}$, that we obtain from numerically optimizing $\omega_D$ such that full oscillations between $\ket{101}$ and $\ket{200}$ are driven by $H_{3\text{\,level}}$ it predicts a drive frequency that only deviates by $|\omega_D - \omega_{\text{opt}}|/\omega_{\text{opt}} = 0.19\%$ for a 200\,ns gate ($0.16\%$ if we include second-order corrections in $\mathcal H_{\text{eff}}$). Doing the same comparison for the gate time, we find deviations of $14\%$ ($5.9\%$ if we include second-order corrections in $\mathcal H_{\text{eff}}$).

While these calculations provide valuable qualitative and quantitative insight into gate operations for long gate times, for faster gates, as considered in our numerical simulations, their accuracy decreases since the truncation of the perturbative expansion in \autoref{eqn:Heff2ndord}, Appendix \ref{sec:calculations}, becomes less accurate.

\subsection{Numerical Analysis}
\label{sec:num_analysis}
For the numerical simulation, we modeled the time-dependent Hamiltonian given in \autoref{eqn:Hamiltonian}, where the parametric drive is implemented analogously to \autoref{eqn:phiext} via flux (phase) modulation,
\begin{equation}
\phi_{\rm{ext},c} = \pi + g(t) \cdot \phi_{\rm{AC}} \cos(\omega_{D} t) \ .
\end{equation}
To minimize leakage errors at the end of the gate, we choose  a flattop-gaussian pulse shape $g(t)$ as described in \autoref{eqn:flattop_gaussian}. The drive frequency  $\omega_D$, modulation amplitude  $\phi_{\rm{AC}}$, the ramp time $\tau$ and a small perturbation around the final gate time $t_{\text{gate}}$ of the envelope $g(t)$ are then optimized. This was done by minimizing analogously to \autoref{eqn:infidelity} the formula \cite{FH, FJ}
\begin{align}
    \label{eqn:infidelity2}
    \epsilon = 1 - \frac{ |\text{Tr} (U^\dagger U_{CZ})|^2 + |\text{Tr} (U^\dagger U)|}{\text{dim}(U)\cdot (\text{dim}(U)+1)}
\end{align}
using the Nelder-Mead algorithm with stochastic initial parameters to achieve high-fidelity gate performance.

\autoref{fig:infidelities}b presents the expected gate infidelities as a function of gate time for the optimized CZ-gate pulses in the light brown line. The gate infidelity is thereby given by \autoref{eqn:infidelity2}
where $U$ represents the process unitary obtained from the simulated time evolution generated by the Hamiltonian in \autoref{eqn:Hamiltonian} and $U_{CZ}$ is the targeted unitary of the exact gate. 
To account for errors due to decoherence, we estimate the decay rate of each quantum state based on its respective  $T_1$  time. These rates are then weighted by the state’s average population during the gate operation, yielding an effective  error rate due to decoherence that is plotted as a dark blue line in \autoref{fig:infidelities}b). The decoherence times used for the simulation are provided in Appendix \ref{sec:parameters}.

To further accelerate the CZ gate between the Fluxonium and the Transmon, we exploit an additional dynamical feature of our system. Periodic driving of the coupler modulates the Floquet energies of the associated eigenstates. At a specific drive amplitude $\phi_{AC}$, this induces an avoided crossing between the Floquet states $|110\rangle_F$ and $|200\rangle_F$. In the vicinity of this resonance, strong effective $ZZ$-interactions are activated, which facilitate the rapid accumulation of a $\pi$-phase on the $|101\rangle$ state.

Our simulation results indicate that for gate durations exceeding $40\,\rm{ns}$, decoherence becomes the dominant source of error. Conversely, for shorter gate times, leakage errors become significant and eventually dominate. This trade-off suggests that an optimal gate fidelity can be achieved with a gate duration of approximately $40\,\rm{ns}$, where the infidelity remains well below  $10^{-3}$.

To mitigate residual leakage-state population at the end of the gate, we introduce a novel gate optimization component and apply a secondary, smaller flux pulse to the coupler \cite{EB}, so that the total flux bias is of the form
\begin{equation}
\phi_{\rm{ext},c} = \pi + g(t) \cdot \phi_{\rm{AC}} \cos(\omega_{D} t) + g_{c}(t) \cdot \phi_{\rm{AC},c} \cos(\omega_{D_c} t)
\end{equation}
The infidelities of gates that are driven by these two-tone drives are shown as a green line in \autoref{fig:infidelities}b) and are significantly lower than those for single tone drives. Our simulations thus confirm that this second correction pulse can reduce closed-loop gate infidelities by up to four orders of magnitude, as also demonstrated for a $30\,\rm{ns}$ gate in \autoref{fig:infidelities}d. Moreover, since this approach effectively suppresses leakage errors even for short gate times, it has the potential to shift the threshold at which leakage becomes relevant. This, in turn, could enable the use of shorter gate times which comes along with a reduction of decoherence-induced errors.

The achievable gate infidelities for two-tone drives as shown in \autoref{fig:infidelities}b) vary between different gate times and do not show a fully conclusive trend, since the optimization might find different local minima in the highly complex landscape in the optimization space. Nonetheless most of the achieved infidelities are orders of magnitude below the errors due to decoherence, so that one can expect significant improvements even for sub-optimal two-tone drives. These observations highlight the potential of employing multiple drive tones for suppressing leakage errors in quantum gates.

In a circuit incorporating a tunable Transmon ancilla, a CZ gate can also be realized by biasing the Transmon to a point where the $\ket{101}$ and $\ket{200}$ states become resonant. While our simulations indicate that high-fidelity CZ-gate implementations are feasible using this approach, potential drawbacks may arise in a larger-scale system. For a more detailed discussion, see Appendix \ref{sec:CZ_alternative}.

%% file: 4_Larger_system.tex
\section{\label{sec:spectator}Analysis of spectator errors}
\begin{figure}[!htp]
    \centering \includegraphics[width=\columnwidth]{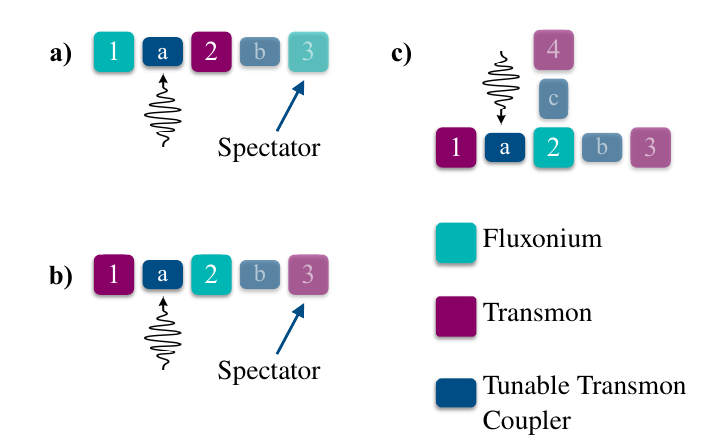}
    \caption{Simulated systems analyzed in this section. a) represents the F-T-F system, while b) illustrates the T-F-T system. Two additional Transmon spectators are coupled to the Fluxonium in c). The parameters for qubits 1 and 2 are the same like those introduced in \autoref{sec:circuit}.}
    \label{fig:spectator}
\end{figure}

To assess the impact of spectator qubits on the gate performance, we analyzed both a Fluxonium - Transmon - Fluxonium (F-T-F) configuration, with an additional Fluxonium as a spectator qubit (cf. \autoref{fig:spectator}a) and a Transmon - Fluxonium - Transmon (T-F-T) configuration with a single and two extra Transmons as a spectator qubits (cf. \autoref{fig:spectator}b and c). 

In these simulations, we implemented a two-qubit gate on the first two qubits, using the parameters of \autoref{sec:circuit} B, while coupling these to a third spectator qubit. The parameters for the third qubit and the coupling element, provided in Appendix \ref{sec:parameters}, were specifically selected to completely suppress pairwise $ZZ$-crosstalk. Nonetheless, when simulating the full system, a small sub-kHz $ZZ$-interaction emerges due to the increased hybridization of the middle qubit. This interaction could be mitigated by fine-tuning the external magnetic flux pulses of the respective couplers, see Appendix \ref{sec:ZZLS} for further details. Yet, to avoid tuning the couplers away from their sweet spot, this was not done here.
Moreover, our simulation shows that $ZZ$-crosstalk between qubits 1 and 3, as well as $ZZZ$-type interactions, defined by 
\begin{align}
    \xi &= E_{111} - E_{100} - E_{010} - E_{001} + 2E_{000} \notag \\
    &- \zeta_{12} - \zeta_{13} - \zeta_{23} \ ,
\end{align}
are extremely small ($<1\,\rm{Hz}$). Here, the couplers are unpopulated and are left out in the subscripts of the energies and $\zeta_{ij}$ denotes the $ZZ$-interactions as defined in \autoref{eqn:zeta} between qubits $i$ and $j$.

\begin{figure}[!htp]
    \centering  \includegraphics[width=\columnwidth]{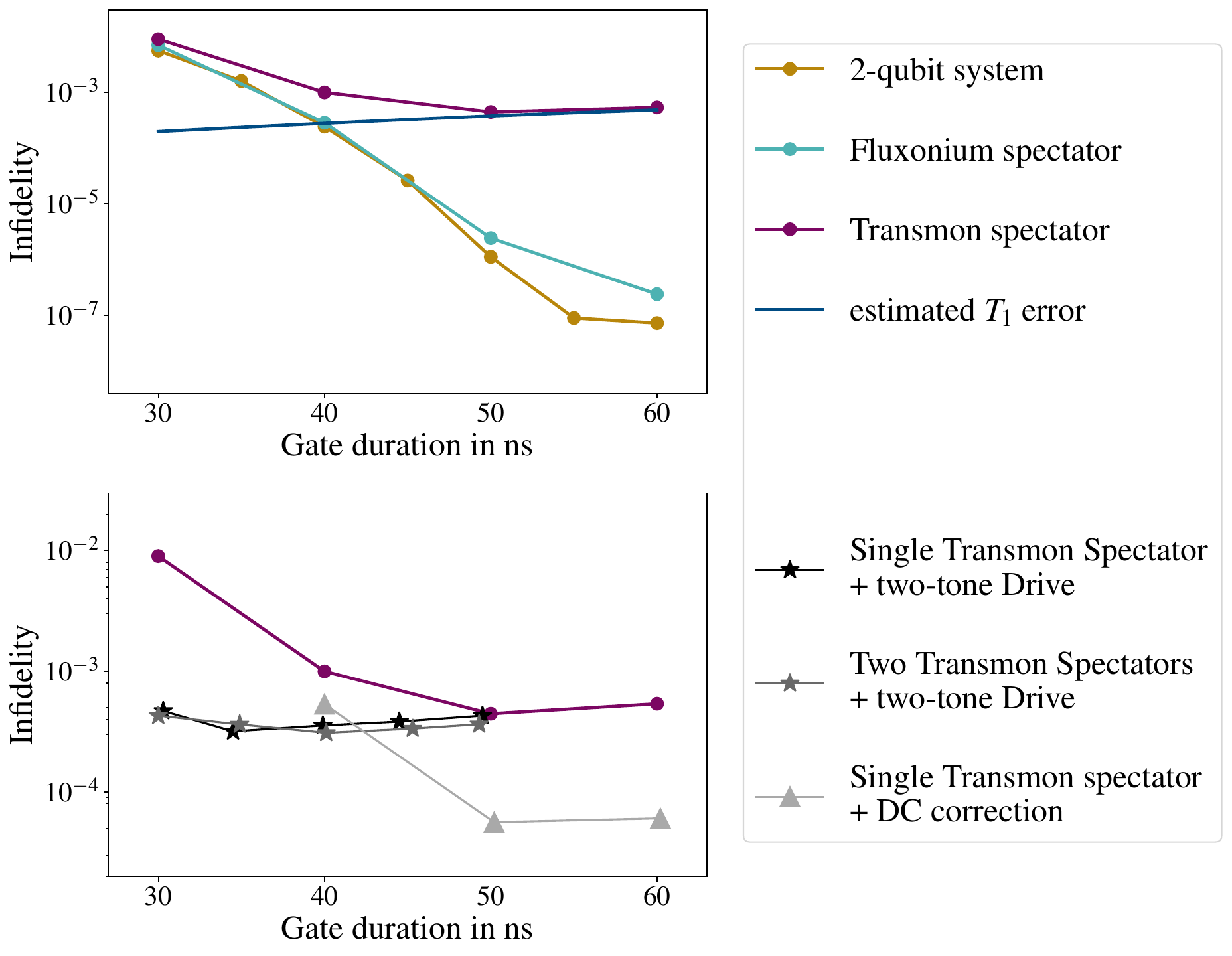}
    \caption{Gate performance of the CZ-gate in the presence of spectator qubits. The light brown line represents the simulated gate infidelity of the isolated 2-qubit system (same light brown line as in \autoref{fig:infidelities} b). The cyan line corresponds to gate infidelities in the presence of a Fluxonium spectator and the purple line shows gate infidelities, when an additional Transmon spectator is coupled to the system. All data represented in the upper panel of the figure correspond to simulations of a single-tone flux drive of the coupler.  
    The lower panel of the figure indicates that the fidelity can be pushed by an order of magnitude in presence of one or two Transmon spectators by applying an additional drive tone to the coupler flux pulse as introduced in \autoref{sec:CZ_gate} B to mitigate additional leakage errors. At longer gate times, where phase errors become dominant, flux-tunable Transmons allow for the application of additional DC pulses to coupler and the spectator, resulting in an improvement in fidelity by an order of magnitude. } \label{fig:spectator_simulation}
\end{figure}

For both the F-T-F and the T-F-T system, we optimize the parameters of the single-tone flux drive to achieve the optimial 2-qubit gate fidelity between the first two qubits, where we calculate the expected closed-system two-qubit error using the formula of \autoref{eqn:infidelity2}. To obtain the corresponding two-qubit unitary, we trace out the spectator qubit from the resulting $8\times8$ process matrix, obtained by simulating the dynamics generated by the full time-dependent Hamiltonian of the system, and normalize the result by a factor $1/2$.

\autoref{fig:spectator_simulation} illustrates the influence of the spectator qubit on the gate performance. By comparing the simulated gate infidelity of an isolated two-qubit system (light brown line on the upper plot) with that of a system coupled to either a Fluxonium spectator (cyan line) or a Transmon spectator (purple line), we observe that for short gate times, where our gate scheme is not limited by errors due to decoherence, the infidelity remains within the same order of magnitude. All leakage state populations are suppressed well at the end of the gate for the Fluxonium spectator, despite the coupling to the additional spectator qubit. For the Transmon spectator we observe a slightly increased risk for leakage state population, particularly for short gate times. 

We find that these leakage state population can be mitigated via a similar strategy as described in \autoref{sec:CZ_gate}. To this end, we apply additional ac flux-pulses as correction pulses to coupler $a$. As indicated by the black stars in the lower panel of \autoref{fig:spectator_simulation}, these decrease the infidelity to values below the coherence limit and very close to the infidelity obtained in an isolated 2-qubit system. Introducing a second Transmon spectator into the system shows that there is no increased risk of additional leakage. In fact, the simulated gate infidelities in this system are sometimes even slightly lower, potentially due to the convergence to a more favorable local minimum in the numerical optimization.

\begin{figure}[!htp]
    \centering \includegraphics[width=\columnwidth]{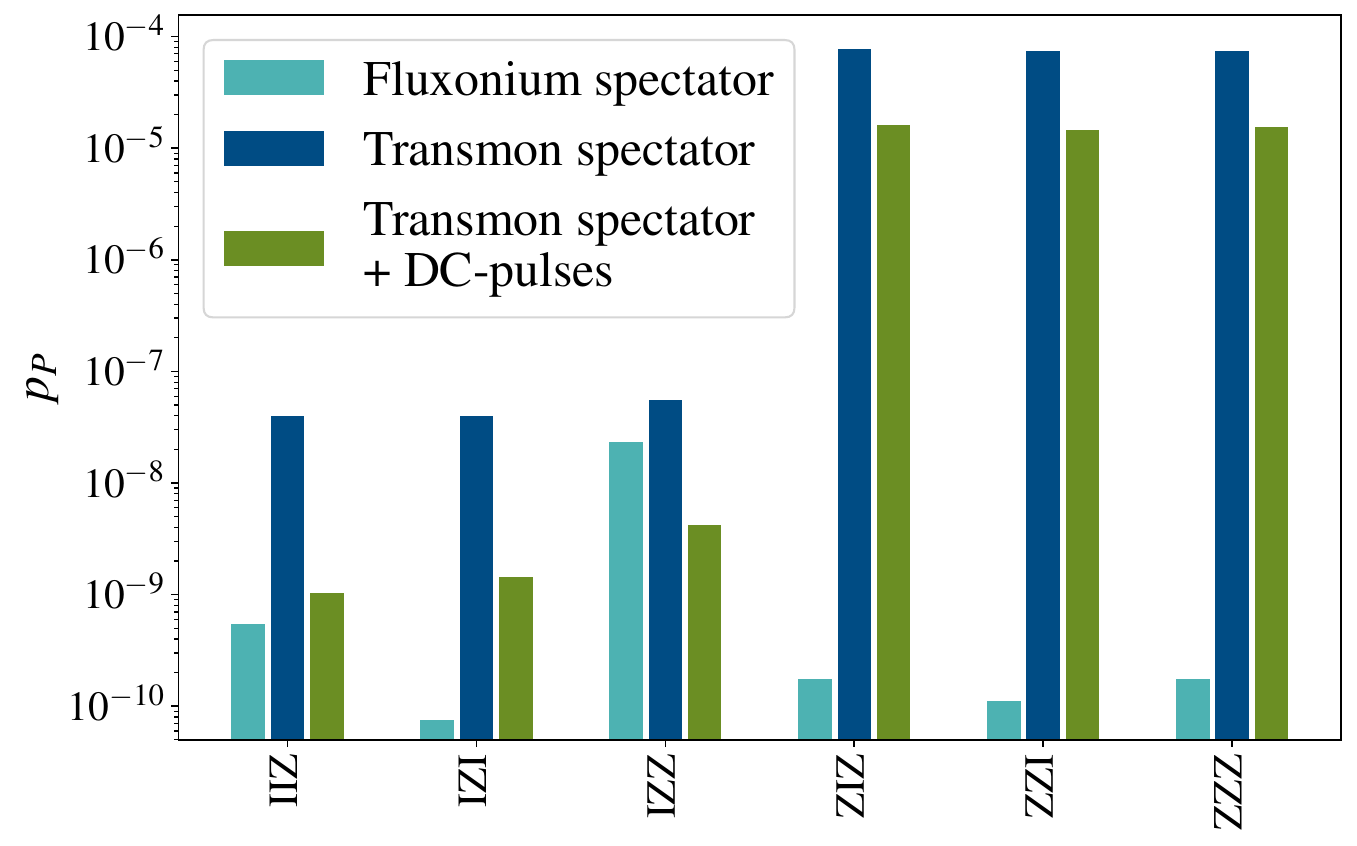}
    \caption{Dominant Pauli error contributions, computed using \autoref{eqn:pauli_weights} for a 50\,ns gate. Dark blue bars correspond to the T-F-T system, while cyan bars represent the F-T-F configuration. Green bars show the results after optimizing the T-F-T system with an additional dc flux pulse applied to coupler b and the spectator Transmon, demonstrating a reduction of phase errors.}
    \label{fig:pauli_weights}
\end{figure}

For longer gate durations, the infidelity does not continue to decrease as one might expect from the isolated two-qubit system. Instead, it reaches a lower bound and then begins to increase again. This effect is particularly pronounced in the presence of a Transmon spectator, where gate fidelities saturates around $10^{-4}$. Since the residual errors visible in these simulations must correspond to Pauli-type errors, we compute their probabilities by 
\begin{align}
    p_P &= \left| \frac{ \text{Tr} \bigl(U_\text{err}^\dagger \cdot \mathcal{P} \bigr)}{8} \right|^2 \ ,
    \label{eqn:pauli_weights}
\end{align}
where $U_{\text{err}} = U^\dagger \cdot (U_{CZ} \otimes \displaystyle \mathds{1})$, $U$ is the simulated $8\times8$ process matrix, and $\mathcal P$ are tensor products of Pauli matrices. The dominant contributions to the infidelity are phase errors, which we plot on a logarithmic scale for the considered configurations in \autoref{fig:pauli_weights}. These are significantly larger in the system with a Transmon spectator (dark blue bars), compared to the setup with a Fluxonium spectator (cyan bars). This difference most likely arises from strong interactions in the second excited state, which are not tuned to zero by our coupler model \cite{BV}. These interactions lead to dispersive shifts and, consequently, phase errors when the second excited state of the Fluxonium is populated during the gate. 

A possible strategy to mitigate these phase errors is to assume that the Transmons in the system are flux-tunable. We can then apply a dc flux pulse to the coupler labeled $b$ in \autoref{fig:spectator}, as well as to the spectator Transmon. Optimizing a 50\,ns gate in this setup reveals that the fidelity can be improved by an order of magnitude, reaching an error of $5.6 \times 10^{-5}$ (see grey triangle in \autoref{fig:spectator_simulation}). As indicated by the green bars in \autoref{fig:pauli_weights}, the probability of a phase error decreases by at least an order of magnitude compared to the setup without tunable Transmons.

Another strategy to mitigate additional crosstalk is to use dynamical decoupling schemes, which decouple spectator qubits when interactions are unwanted. Such techniques have already been successfully implemented in superconducting qubit platforms (e.g., \cite{EF}).

While further means to reduce the influence of spectator qubits should be explored in future research, our results suggest very promising properties for fast gate operations. In this regime, both decoherence and spectator-induced phase errors are significantly suppressed. In particular, for a $40\,\rm{ns}$ gate, the estimated decoherence-induced infidelity is similar to that of the closed-system infidelity in the presence of either a Fluxonium or a Transmon spectator, suggesting that the gate remains robust and high-fidelity even in an integrated architecture.

%% file: 5_Conclusions.tex
\section{\label{sec:conclusions}Conclusions}
To summarize, we have proposed a hybrid quantum architecture combining Fluxonium and Transmon qubits and demonstrated its strong scalability potential. Key advantages include parameter regimes that mitigate the limitations of capacitive loading, which is often a challenge in Fluxonium systems, and a reduced risk of frequency crowding, achieved through the alternating arrangement of different qubit types. Additionally, the system exhibits highly localized qubit states, an absence of $ZZ$-crosstalk in idle conditions, and a high-fidelity CZ-gate that remains robust even in the presence of spectator qubits. Notably, the CZ-gate operates without requiring tuning of qubit frequencies, which eliminates the need to move qubits across resonances, a common source of spectator-induced errors, thereby further enhancing its scalability.

Interesting steps of future research are, besides an experimental implementation of the architecture and gate, more in depth studies of avenues to exploit the promising aspects of the architecture for efficient and low resource quantum error correction implementations.

\begin{acknowledgments}
This work has received funding from the German Federal Ministry of Education and Research via the funding program quantum technologies - from basic research to the market under contract numbers 13N15684 "GeQCoS" and 13N16182 "MUNIQC-SC". It is also part of the Munich Quantum Valley, which is supported by the Bavarian state government with funds from the Hightech Agenda Bayern Plus.
\end{acknowledgments}

%% file: Appendix.tex
\section{\label{sec:circuitderivation} Circuit Hamiltonian derivation}
The Hamiltonian for the circuit can be derived performing a Legendre transformation of the corresponding Lagrangian, which is given by
\begin{equation}
L = \frac{1}{2} \mathbf{\dot{\Phi}}^T \mathbf{C} \mathbf{\dot{\Phi}} - V(\mathbf{\Phi}) \ ,
\end{equation}

where $\mathbf{\Phi} = (\Phi_1, \Phi_{c_1}, \Phi_{c_2}, \Phi_2)^T$ is a vector containing the node fluxes of the circuit shown in \autoref{fig:circuit2}. The capacitance matrix $\mathbf{C}$ is defined as:
\begin{equation}
\mathbf{C} = \begin{pmatrix}
C_1+C_{1c} & -C_{1c} & 0 & 0 \\
-C_{1c} & C_c + C_g +C_{1c} & -C_c & 0 \\
0 & -C_c & C_c + C_g + C_{2c} & -C_{2c} \\
0 & 0 & -C_{2c} & C_2
\end{pmatrix}
\end{equation}
and the potential can be written as 
\begin{align}
    V(\mathbf{\Phi}) &= \frac{1}{2}E_{L,1} \left(\frac{\Phi_1-\Phi_{\text{ext},1}}{2\pi\Phi_0}\right)^2 - E_{J,1} \cos \frac{\Phi_1}{2\pi\Phi_0} \notag \\
    &- E_{J,c_u} \cos \left( \frac{\Phi_{c1} - \Phi_{c2}}{2\pi\Phi_0}\right) \notag \\
    &- E_{J,c_l} \cos \left( \frac{\Phi_{c2} - \Phi_{c1} - \Phi_{\text{ext},c}}{2\pi\Phi_0}\right) \notag \\ 
    & - E_{J,2} \cos \frac{\Phi_2}{2\pi\Phi_0}\ ,
\end{align} 
where $\Phi_0$ denotes the magnetic flux quantum and we are using the reduced flux variables $\phi_i = \Phi_i/(2\pi\Phi_0)$ in the following. We can derive the Hamiltonian by performing a Legendre transformation. The canonical conjugate charges are given by $\mathbf{q} = \mathbf{C}^{-1}\mathbf{\dot{\Phi}}$. Renormalizing the charges by a factor of $1/2e$ we can express the system in terms of the fluxes and the Cooper pair number operators $n_i$
\begin{align}
    H = 4\cdot\mathbf{n}^T \mathbf{E_C} \mathbf{n} + V(\mathbf{\Phi}) \ ,
\end{align}
where we defined the matrix $\mathbf{E_C} = \mathbf{C}^{-1} \cdot \frac{e^2}{2h}$. In order to quantize the Hamiltonian, we note, that the potential $V(\mathbf{\Phi})$ is dependent only on the term $\Phi_{c1} - \Phi_{c2}$. Therefore we change the basis according to 
\begin{align}
    \Phi_\pm &= \Phi_{c1} \pm \Phi_{c2} \\
    n\pm &= \frac{1}{2}(n_{c1} \pm n_{c2}) \ ,
\end{align}
where we have introduced an additional factor $1/2$, to make sure, that after quantization, the commutator $[n_i, q_j] = i\delta_{ij}$ still holds in the plus-minus basis. Since $\phi_+$ does not appear in the potential, the conjugate charge $n_+$ is a conserved quantity. Fixing it to its ground-state value, the capacitive coupling terms vanish, such that the +-mode decouples and is left out in the following. From now on, we refer to the $-$-mode as the coupler mode.

Each mode is quantized separately using a harmonic oscillator. For the two Transmon mode, we expand the cosine potential up to the fourth order and define the charge and flux operators as,
\begin{align}
    n_2 &= \frac{i}{\sqrt{2}} \left( \frac{E_{J,2}}{8E_{C,22}}\right)^{1/4} \left( a_2^\dagger - a_2 \right) \\ 
    \phi_2 &= \frac{1}{\sqrt{2}} \left( \frac{8E_{C,22}}{E_{J,2}}\right)^{1/4} \left( a_2^\dagger + a_2 \right), 
\end{align}
where $a^\dagger$ and $a$ are the bosonic creation and annihilation operators, satisfying the commutation relation \mbox{$[a, a^\dagger] = 1$}. For the coupler, we set the external flux to $\pi$ and can then analogously define the creation and annihilation operators, 
\begin{align}
    n_c &= \frac{i}{\sqrt{2}} \left( \frac{E_{J_\Delta}}{8E_{C,cc}}\right)^{1/4} \left( a_c^\dagger - a_c \right) \\ 
\phi_c &= \frac{1}{\sqrt{2}} \left( \frac{8E_{C,cc}}{E_{J_\Delta}}\right)^{1/4} \left( a_c^\dagger + a_c \right) \ ,
\end{align}
where $E_{J_\Delta}$ denotes the difference of the two Josephson junctions of the coupler $E_{J_\Delta} = |E_{J,c_u} - E_{J,c_l}|$. For the Fluxonium, we define the charge and flux operators as,
\begin{align}
n &= \frac{i}{\sqrt{2}} \left( \frac{E_L}{8E_C}\right)^{1/4} \left( a^\dagger - a \right) \\
\phi &= \frac{1}{\sqrt{2}} \left( \frac{8E_C}{E_L}\right)^{1/4} \left( a^\dagger + a \right)
\end{align}
It is important to note that the single Fluxonium Hamiltonian must be simulated with a large number of levels (we used 100) until the lowest energy levels and the respective transition matrix elements converge. After testing the convergence, we kept the 5 lowest lying states and energies for simulating the whole FTT-architecture. Using these equations we arrive at \autoref{eqn:quantH}.

\section{\label{sec:parameters}Parameter sets for simulation}
\begin{figure}[!htp]
    \centering
    \includegraphics[width=\linewidth]{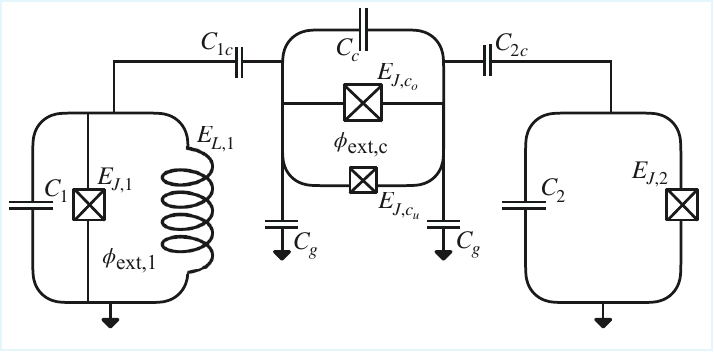}
    \caption{Lumped element circuit investigated in \autoref{sec:circuit} and \autoref{sec:CZ_gate}}
    \label{fig:circuit2}
\end{figure}
The circuit with all parameters labelled is drawn in \autoref{fig:circuit2}. We use the parameter sets of \autoref{tab:params0} for our simulations in \autoref{sec:CZ_gate} and the parameters of \autoref{tab:params1} and \autoref{tab:params2} for our simulations in \autoref{sec:spectator}. In this section, we denote the indices for the coupler parameters with $a$ and $b$, where coupler $a$ couples the first and second qubits, and coupler $b$ couples the second and third qubits, respectively. These values for the coupling capacitances have been confirmed to avoid the problem of capacitive loading.

\begin{table}[!htp]
\centering
\begin{tabular*}{0.9\columnwidth}{@{\extracolsep{\fill}} lllllll}
\toprule \\[-0.3cm]
$C_1$ & $C_2$ & $C_c$ & $C_g$ & $C_{1c}$ & $C_{2c}$ & \\ \midrule
18 & 87.8 & 22 & 38 & 6 & 15.5 & in fF \\[0.3cm]
\end{tabular*}
\begin{tabular*}{0.9\columnwidth}{@{\extracolsep{\fill}} llllll}
\\[-0.24cm]
$E_{L,1}$ & $E_{J,1}$ & $E_{J,c_o}$ & $E_{J,c_u}$ & $E_{J,2}$  & \\ \midrule
1.6 & 6.1 & 12.822 & 7.5 & 13.6 & in GHz \\[0.05cm]
\bottomrule
\end{tabular*}
\caption{Parameter set for the simulation in \autoref{sec:CZ_gate}}
\label{tab:params0}
\end{table}
For the estimation of the errors due to decoherence we considered the following decay rates: 
\begin{align*}
    \Gamma_{100 \rightarrow 000} &= \phantom{2}4 \, \mathrm{kHz} \qquad \qquad \Gamma_{200\rightarrow100} = 22 \, \mathrm{kHz} \\
    \Gamma_{001 \rightarrow 000} &= 25 \, \mathrm{kHz} \qquad \qquad \Gamma_{010\rightarrow 000} = 40 \, \mathrm{kHz}
\end{align*}

\begin{table}[!htp]
\centering
\begin{tabular*}{0.9\columnwidth}{@{\extracolsep{\fill}} llllll}
\toprule \\[-0.3cm]
$C_3$ & $C_b$ & $C_{g,b}$ & $C_{2b}$ & $C_{3b}$ &  \\ \midrule
21 & 25 & 32.5 & 15 & 6.5 & in fF \\[0.3cm]
\end{tabular*}
\begin{tabular*}{0.9\columnwidth}{@{\extracolsep{\fill}} lllll}
\\[-0.24cm]
$E_{L,3}$ & $E_{J,3}$ & $E_{J,b_o}$ & $E_{J,b_u}$ & \\ \midrule
1.96 & 6.2 & 13.11 & 7.5 & in GHz \\[0.05cm]
\bottomrule
\end{tabular*}
\caption{Parameter set for F-T-F simulation in \autoref{sec:spectator}}
\label{tab:params1}
\end{table}

\begin{table}[!htp]
\centering
\begin{tabular*}{0.9\columnwidth}{@{\extracolsep{\fill}} llllll}
\toprule \\[-0.3cm]
$C_3$ & $C_b$ & $C_{g,b}$ & $C_{2b}$ & $C_{3b}$ & \\ \midrule
90 & 30 & 35 & 6 & 14 & in fF \\[0.3cm]
\end{tabular*}
\begin{tabular*}{0.9\columnwidth}{@{\extracolsep{\fill}} llllll}
\\[-0.24cm]
$C_4$ & $C_c$ & $C_{g,c}$ & $C_{2c}$ & $C_{4c}$ \\ \midrule
93 & 26 & 34 & 4.8 & 16 & in fF \\[0.3cm]
\end{tabular*}
\begin{tabular*}{0.9\columnwidth}{@{\extracolsep{\fill}} ccccccc}
\\[-0.24cm]
$E_{J,3}$ & $E_{J,b_o}$ & $E_{J,b_u}$ & $E_{J,4}$ & $E_{J,c_o}$ & $E_{J,c_u}$ & \\ \midrule
15.1 & 13.471 & 6.5 & 16.7 & 14.524 & 7.8 & in GHz \\[0.05cm]
\bottomrule
\end{tabular*}
\caption{Parameter set for the single and two Transmon simulation in \autoref{sec:spectator}}
\label{tab:params2}
\end{table}
We demonstrate the feasibility of these parameters for scaling the qubit system by simulating a CZ gate using the alternative parameter sets. In both cases, the energy of the $|110\rangle$ state exceeds that of the $|200\rangle$ state. To exploit the strong $ZZ$-interactions at the avoided crossing of the respective Floquet states, we select a parameter regime where the coupler is biased at its upper sweet spot. By driving the coupler flux near the resonance condition for the $|101\rangle \leftrightarrow |200\rangle$ transition, we achieve an optimal CZ gate within $52\,\rm{ns}$ and with an infidelity of $2.07 \times 10^{-6}$ for the alternative Fluxonium parameters, and $65\,\rm{ns}$ with an infidelity of $2.49 \times 10^{-4}$ for the alternative Transmon parameters.

\FloatBarrier

\section{\label{sec:Cloading}Capacitive loading}
We aim to avoid the capacitive-loading problem, a common challenge when integrating Fluxoniums into a square-lattice architecture. Since the charging energy $E_C$ of Fluxoniums is significantly larger than that of Transmons, the physical area of the corresponding smaller Fluxonium shunt capacitance becomes the limiting factor, when coupling additional qubits. To demonstrate that, within our parameter regime, we remain within the available capacitance budget, we compare a Fluxonium coupled to a single additional qubit with one coupled to two additional qubits. The capacitance matrices of the two architectures can be written as
\begin{align}
    \mathbf{C} &= \begin{pmatrix}
        C_1+C_c & -C_c \\
        -C_c & C + C_c
    \end{pmatrix} \\ \mathbf{\tilde C} &= \begin{pmatrix}
        \tilde C_1 + 2C_c & -C_c & -C_c \\
        -C_c & C+C_c & 0 \\ 
        -C_c & 0 & C+C_c
    \end{pmatrix}
\end{align}
We require the quantity $E_C \propto \mathbf{C}^{-1}_{[1,1]}$ of the Fluxonium to remain the same in both architectures. We therefore compute the corresponding matrix element of the inverse capacitance matrices and expand it in powers of $C_c$, assuming the coupling capacitance is small compared with the shunt capacitances $C_1$ and $C$:
\begin{align}
    \mathbf{C}^{-1}_{[1, 1]} &= \frac{1}{C_1} - \frac{C_c}{C_1^2} + \mathcal O(C_c^2) \\
    \mathbf{\tilde C}^{-1}_{[1, 1]} &= \frac{1}{\tilde C_1} - \frac{2C_c}{\tilde C_1^2} + \mathcal O(C_c^2) \ .
\end{align}
Setting $\mathbf{C}^{-1}_{[1,1]} = \mathbf{\tilde C}^{-1}_{[1,1]}$ and solving for the shunt capacitance $\tilde C_1$, we find that, to leading order,
\begin{align}
    \tilde C_1 \approx C_1 - C_c \ .
\end{align}
It follows that $C_1>3C_c$ must hold for a Fluxonium that is coupled to only a single additional qubit, in order to allow for three further neighboring qubits in a scaled architecture. Using the parameter regime of Appendix \ref{sec:parameters}, we conclude that it is feasible to couple three additional Transmons to the Fluxonium with comparable coupling strengths.

\section{\label{sec:ZZcrosstalk}$ZZ$-Crosstalk}
We can calculate the energies of \autoref{eqn:zeta} using time-independent perturbation theory. Since the direct coupling between the qubits is much smaller than the coupling between the qubits and the coupler, relevant terms appear up to fourth order: 
\begin{align}
    \label{eqn:zeta2}
    \zeta^{(2)} =& \frac{g_{101\rightarrow 200}^2}{E_{101}^{(0)}-E_{200}^{(0)}} + \frac{g_{101\rightarrow 002}^2}{E_{101}^{(0)}-E_{002}^{(0)}} \\ 
    \zeta^{(3)} =& \frac{2g_{101\rightarrow 002}g_{110\rightarrow 200}g_{101\rightarrow 110}}{(E_{101}^{(0)}-E_{200}^{(0)})(E_{101}^{(0)}-E_{110}^{(0)})} \notag \\
    &+ \frac{2g_{101\rightarrow 002}g_{011\rightarrow 002}g_{101\rightarrow 001}}{(E_{101}^{(0)}-E_{002}^{(0)})(E_{101}^{(0)}-E_{011}^{(0)})} \notag \\ 
    &+ \frac{2g_{101\rightarrow 110}g_{110\rightarrow 011}g_{101\rightarrow 011}}{(E_{101}^{(0)}-E_{110}^{(0)})(E_{101}^{(0)}-E_{011}^{(0)})} \notag \\
    &- \frac{2g_{100\rightarrow 010}g_{001\rightarrow 010}g_{100\rightarrow 001}}{(E_{100}^{(0)}-E_{010}^{(0)})(E_{100}^{(0)}-E_{001}^{(0)})} \notag \\
    &- \frac{2g_{100\rightarrow 010}g_{001\rightarrow 010}g_{100\rightarrow 001}}{(E_{001}^{(0)}-E_{010}^{(0)})(E_{001}^{(0)}-E_{100}^{(0)})} \\ 
    \zeta^{(4)} =& \frac{g_{101\rightarrow 110}^2g_{110\rightarrow 200}^2}{(E_{101}^{(0)}-E_{110}^{(0)})^2(E_{101}^{(0)}-E_{200}^{(0)})} \notag \\
    &+ \frac{g_{101\rightarrow 011}^2g_{011\rightarrow 002}^2}{(E_{101}^{(0)}-E_{011}^{(0)})^2(E_{101}^{(0)}-E_{002}^{(0)})} \notag \\ 
    &+ \frac{g_{101\rightarrow 110}^2g_{110\rightarrow 020}^2}{(E_{101}^{(0)}-E_{110}^{(0)})^2(E_{101}^{(0)}-E_{020}^{(0)})} \notag \\ 
    &+ \frac{g_{101\rightarrow 011}^2g_{011\rightarrow 020}^2}{(E_{101}^{(0)}-E_{011}^{(0)})^2(E_{101}^{(0)}-E_{020}^{(0)})} \notag \\ 
    &+ \frac{2g_{101\rightarrow 110}g_{110\rightarrow 020}g_{011\rightarrow 020}g_{101\rightarrow 011}}{(E_{101}^{(0)}-E_{110}^{(0)})(E_{101}^{(0)}-E_{020}^{(0)})} \notag \\ 
    &\times \frac{1}{(E_{101}^{(0)}-E_{011}^{(0)})} \notag 
\end{align}
\begin{align}
    &- \frac{g_{101\rightarrow 110}^2g_{101\rightarrow 011}^2}{(E_{101}^{(0)}-E_{110}^{(0)})^2(E_{101}^{(0)}-E_{011}^{(0)})} \notag \\
    &- \frac{g_{101\rightarrow 110}^2g_{101\rightarrow 011}^2}{(E_{101}^{(0)}-E_{011}^{(0)})^2(E_{101}^{(0)}-E_{110}^{(0)})} \notag \\
    &- \frac{g_{100\rightarrow 010}^2g_{001\rightarrow 010}^2}{(E_{100}^{(0)}-E_{010}^{(0)})^2(E_{100}^{(0)}-E_{001}^{(0)})} \notag \\ 
    &- \frac{g_{100\rightarrow 010}^2g_{001\rightarrow 010}^2}{(E_{001}^{(0)}-E_{010}^{(0)})^2(E_{001}^{(0)}-E_{100}^{(0)})}
    \label{eqn:zeta4}
\end{align}
Considering the different Energy structure of the Fluxonium compared to the Transmon, we observe, that $|E_{101}-E_{200}| \gg |E_{101}-E_{002}|$ as well as $|E_{101}-E_{110}| \gg |E_{101}-E_{011}|$. Therefore, only the first term of each, $\zeta^{(2)}$, $\zeta^{(3)}$ and $\zeta^{(4)}$ contributes significantly. While $\zeta^{(2)}$ and $\zeta^{(4)}$ are positive, $\zeta^{(3)}$ is negative due to the different sign of the couplings, connecting the Transmon and the Fluxonium to the coupler.
\begin{figure}[!htp]
    \centering
    \includegraphics[width=\columnwidth]{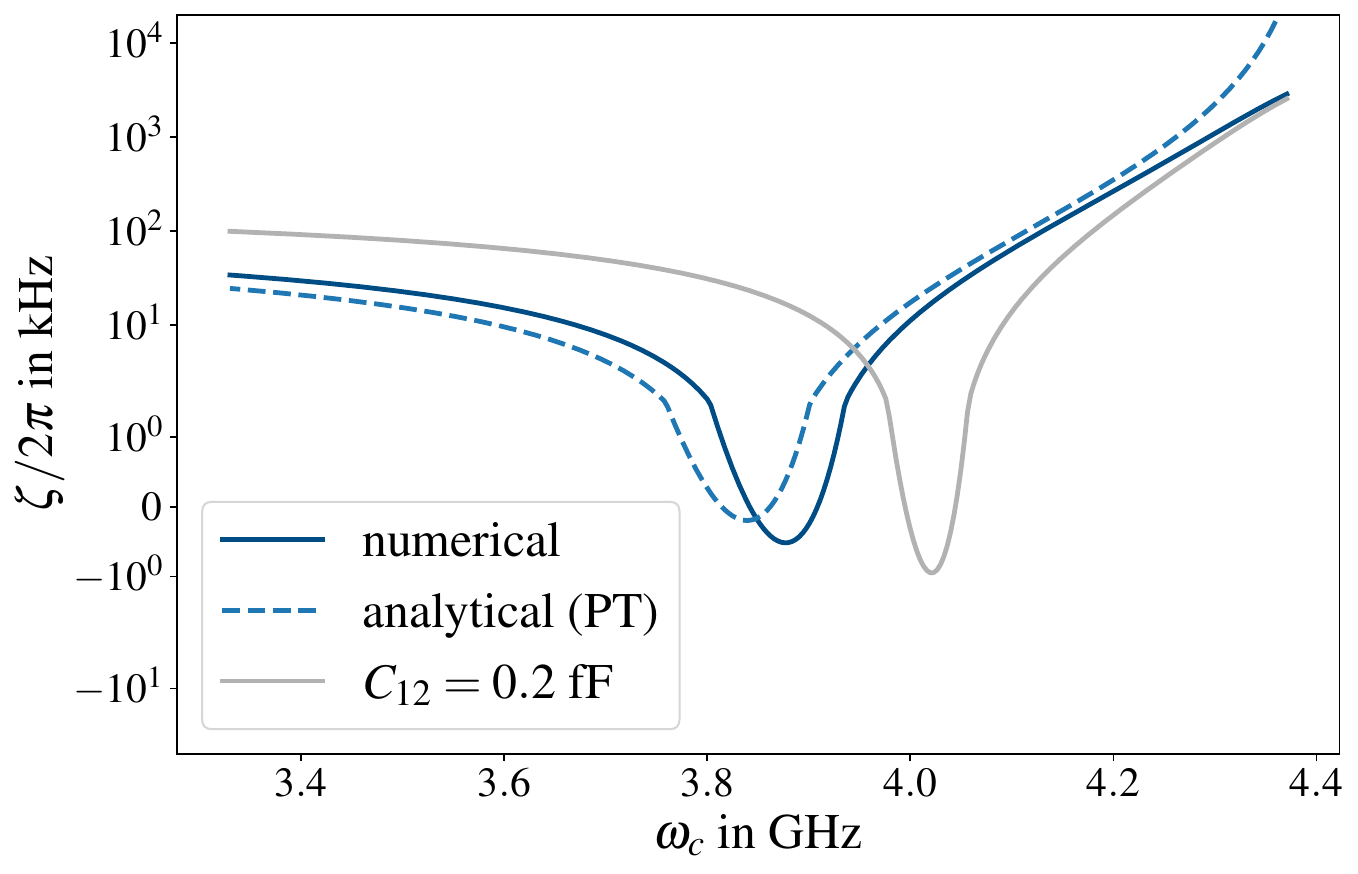}
    \caption{$ZZ$-crosstalk $\zeta/2\pi$ versus the coupler frequency $\omega_d$. The dashed line is calculated by perturbation theory, while the solid blue line is calculated from a numerical diagonalisation of the truncated Hamiltonian in \autoref{eqn:quantH}. The deviation is due to counter-rotating wave terms, that are neglected in the analytical calculation. The gray line shows, that even in presence of a direct parasitic capacitance between the two qubits, $ZZ$-crosstalk can still be eliminated by readjusting the coupler frequency.}
    \label{fig:zetaPT}
\end{figure}
\autoref{fig:zetaPT} shows, that perturbation theory is in good agreement with the result obtained from direct diagonalisation of the Hamiltonian. The deviations are of the expected order and arise from counter-rotating wave terms, that are neglected in equations \ref{eqn:zeta2} to \ref{eqn:zeta4}. Furthermore, we show, that additional parasitic capacitances are not detrimental to parasitic interactions, as $ZZ$-crosstalk can be eliminated by readjusting the coupler frequency.

\section{\label{sec:calculations}Analytical calculations}
Considering the drive Hamiltonian from \autoref{eqn:H_D}, we can insert the Ansatz for the pulse $\phi_{\text{ext}}$ from \autoref{eqn:phiext} and apply a Jacobi-Anger expansion 
\begin{align}
    &\sqrt{\cos^2 \phi_{\text{ext}} + d^2 \sin^2 \phi_{\text{ext}}} \\
    =& \sqrt{1-\cos^2(\phi_{AC}\cos(\omega_D t)) + d^2\cos^2(\phi_{AC} \cos (\omega_D t))} \\
    \approx& \sqrt{1+(d^2-1)\left( J_0^2(\phi_{AC}) - 4J_0(\phi_{AC}) J_2(\phi_{AC}) \cos(2\omega_D t)\right)} \\ 
    \approx& \sqrt{1+(d^2-1)J_0^2(\phi_{AC})} \notag \\
    &+ \frac{2(1-d^2)J_0(\phi_{AC})J_2(\phi_{AC})}{\sqrt{1+(d^2-1)J_0^2(\phi_{AC})}} \cos (2\omega_D t) \ ,
\end{align}
where $J_k$ denote the Bessel functions of $k-th$ kind. From this we can read off the coefficients $\alpha$ and $\beta$ introduced in \autoref{sec:CZ_gate} A:
\begin{align}
    \alpha &= E_{J_\Sigma} \sqrt{1+(d^2-1)J_0^2(\phi_{AC})} - E_{J_\Delta} \\
    \beta &= E_{J_\Sigma} \frac{2(1-d^2)J_0(\phi_{AC})J_2(\phi_{AC})}{\sqrt{1+(d^2-1)J_0^2(\phi_{AC})}} \ .
\end{align}
The elements of the matrix \textbf{A} in \autoref{eqn:3lvlH} can be computed using a time-independent perturbation theory. To do this, we expand the cosine of the flux variable $\phi_c$ up to fourth order, which is justified for small zero-point fluctuations $\phi_{\text{ZPF}}$, and apply a rotating wave approximation. By deriving the state corrections for the three states $\ket{101}$, $\ket{110}$, and $\ket{200}$ up to leading order, we extract the elements of the matrix \textbf{A}:
\begin{widetext}
\begin{align}
    A_{00} &= \phi_{\text{ZPF}}^2 \left(1-\frac{\phi_{\text{ZPF}}^2}{2} \right) \frac{\frac{g_{101\rightarrow 110}^2}{\left(E_{101}^{(0)}-E_{110}^{(0)}\right)^2}}{1 + \frac{g_{101\rightarrow 110}^2}{\left(E_{101}^{(0)}-E_{110}^{(0)}\right)^2} + \frac{g_{101\rightarrow 200}^2}{\left(E_{101}^{(0)}-E_{200}^{(0)}\right)^2}} \\
    A_{11} &= \phi_{\text{ZPF}}^2 \left(1-\frac{\phi_{\text{ZPF}}^2}{2} \right) \frac{1}{1 + \frac{g_{110\rightarrow 101}^2}{\left(E_{110}^{(0)}-E_{101}^{(0)}\right)^2} + \frac{g_{110\rightarrow 200}^2}{\left(E_{110}^{(0)}-E_{200}^{(0)}\right)^2}} 
\end{align}
\begin{align}
    A_{22} &= \phi_{\text{ZPF}}^2 \left(1-\frac{\phi_{\text{ZPF}}^2}{2} \right) \frac{\frac{g_{200\rightarrow 110}^2}{\left(E_{200}^{(0)}-E_{110}^{(0)}\right)^2}}{1 + \frac{g_{200\rightarrow 110}^2}{\left(E_{200}^{(0)}-E_{110}^{(0)}\right)^2} + \frac{g_{200\rightarrow 101}^2}{\left(E_{200}^{(0)}-E_{101}^{(0)}\right)^2}} \\
    A_{01} &= \phi_{\text{ZPF}}^2 \left(1-\frac{\phi_{\text{ZPF}}^2}{2} \right) \frac{\frac{g_{101\rightarrow 110}^2}{\left(E_{101}^{(0)}-E_{110}^{(0)}\right)^2}}{\sqrt{1 + \frac{g_{101\rightarrow 110}^2}{\left(E_{101}^{(0)}-E_{110}^{(0)}\right)^2} + \frac{g_{101\rightarrow 200}^2}{\left(E_{101}^{(0)}-E_{200}^{(0)}\right)^2}} \cdot \sqrt{1 + \frac{g_{110\rightarrow 101}^2}{\left(E_{110}^{(0)}-E_{101}^{(0)}\right)^2} + \frac{g_{110\rightarrow 200}^2}{\left(E_{110}^{(0)}-E_{200}^{(0)}\right)^2}}} \\
    A_{12} &= \phi_{\text{ZPF}}^2 \left(1-\frac{\phi_{\text{ZPF}}^2}{2} \right) \frac{\frac{g_{200\rightarrow 110}^2}{\left(E_{200}^{(0)}-E_{110}^{(0)}\right)^2}}{\sqrt{1 + \frac{g_{200\rightarrow 110}^2}{\left(E_{200}^{(0)}-E_{110}^{(0)}\right)^2} + \frac{g_{200\rightarrow 101}^2}{\left(E_{200}^{(0)}-E_{101}^{(0)}\right)^2}} \cdot \sqrt{1 + \frac{g_{110\rightarrow 101}^2}{\left(E_{110}^{(0)}-E_{101}^{(0)}\right)^2} + \frac{g_{110\rightarrow 200}^2}{\left(E_{110}^{(0)}-E_{200}^{(0)}\right)^2}}} \\
    A_{12} &= \phi_{\text{ZPF}}^2 \left(1-\frac{\phi_{\text{ZPF}}^2}{2} \right) \frac{\frac{g_{101\rightarrow 110}}{E_{101}^{(0)}-E_{110}^{(0)}} \cdot \frac{g_{200\rightarrow 110}}{E_{200}^{(0)}-E_{110}^{(0)}}}{\sqrt{1 + \frac{g_{101\rightarrow 110}^2}{\left(E_{101}^{(0)}-E_{110}^{(0)}\right)^2} + \frac{g_{101\rightarrow 200}^2}{\left(E_{101}^{(0)}-E_{200}^{(0)}\right)^2}} \cdot \sqrt{1 + \frac{g_{200\rightarrow 110}^2}{\left(E_{200}^{(0)}-E_{110}^{(0)}\right)^2} + \frac{g_{200\rightarrow 101}^2}{\left(E_{200}^{(0)}-E_{101}^{(0)}\right)^2}}} \\
\end{align}
\end{widetext}
Following \cite{ED, EE} we can extract the time evolution of the three-level system under investigation by
\begin{equation}
    U(t, t_0) = R(t) e^{-i\mathcal K (t)} e^{-i\mathcal H_{\text{eff}} (t-t_0)}e^{i\mathcal K (t_0)} R(t_0) \ ,
\end{equation}
where the operator $R(t) e^{-i \mathcal K(t)}$ describes micromotion, while the effective dynamics is governed by $ \mathcal H_{\text{eff}}$. The perturbative formulas for the Kick-operator as well as the effective Hamiltonian are given in \cite{ED} 
\begin{align}
\label{eqn:Heff2ndord}
    \mathcal H_{\text{eff}} &= \mathcal H^{(0)} + \frac{1}{\omega_D} \sum_{j>0} \frac{1}{j} \left[ \mathcal H^{(j)} , \mathcal H^{(-j)} \right] \notag \\ 
    &+ \frac{1}{\omega_D^2} \sum_{j>0} \frac{1}{j^2} \left[ \Bigr[ \mathcal H^{(j)} , \mathcal H^{(0)} \Bigl], \mathcal H^{(-j)} \right] + h.c. \notag \\ 
    &+ \ \mathcal O\left( \frac{1}{\omega_D^3}\right)
\end{align}
\begin{align}
    \mathcal K(t) &= \frac{1}{i\omega_D} \sum_{j>0} \frac{1}{j} \left( \mathcal H^{(j)}e^{ij\omega_D t} - \mathcal H^{(-j)}e^{-ij\omega_D t} \right) \notag \\ 
    &+ \ \mathcal O\left( \frac{1}{\omega_D^2}\right) \ ,
\end{align}
where the formulas rely on a spectral decomposition of the total Hamiltonian $\mathcal H = \sum_j e^{ij\omega_D t} \mathcal H^{(j)}$. The rotation $R(t)$, that satisfies the condition of a small Hamiltonian compared to the drive frequency $\omega_D$ is for this system given by
\begin{equation}
    R(t) = \exp \left( i \sum_l k_l \omega_D t \ \ket{l}\bra{l} \right) \ ,
\end{equation}
where $k_{101} = 2$, $k_{110} = -1$ and $k_{200} = 0$. 
From $\mathcal H_{\text{eff}}$, we extract the drive frequency $\omega_D$ and the expected gate time $t_{\text{gate}}$. Analytic expressions up to first order can be found in Appendix \ref{sec:calculations}. 

\begin{figure}[!htp]
    \centering
    \includegraphics[width=\columnwidth]{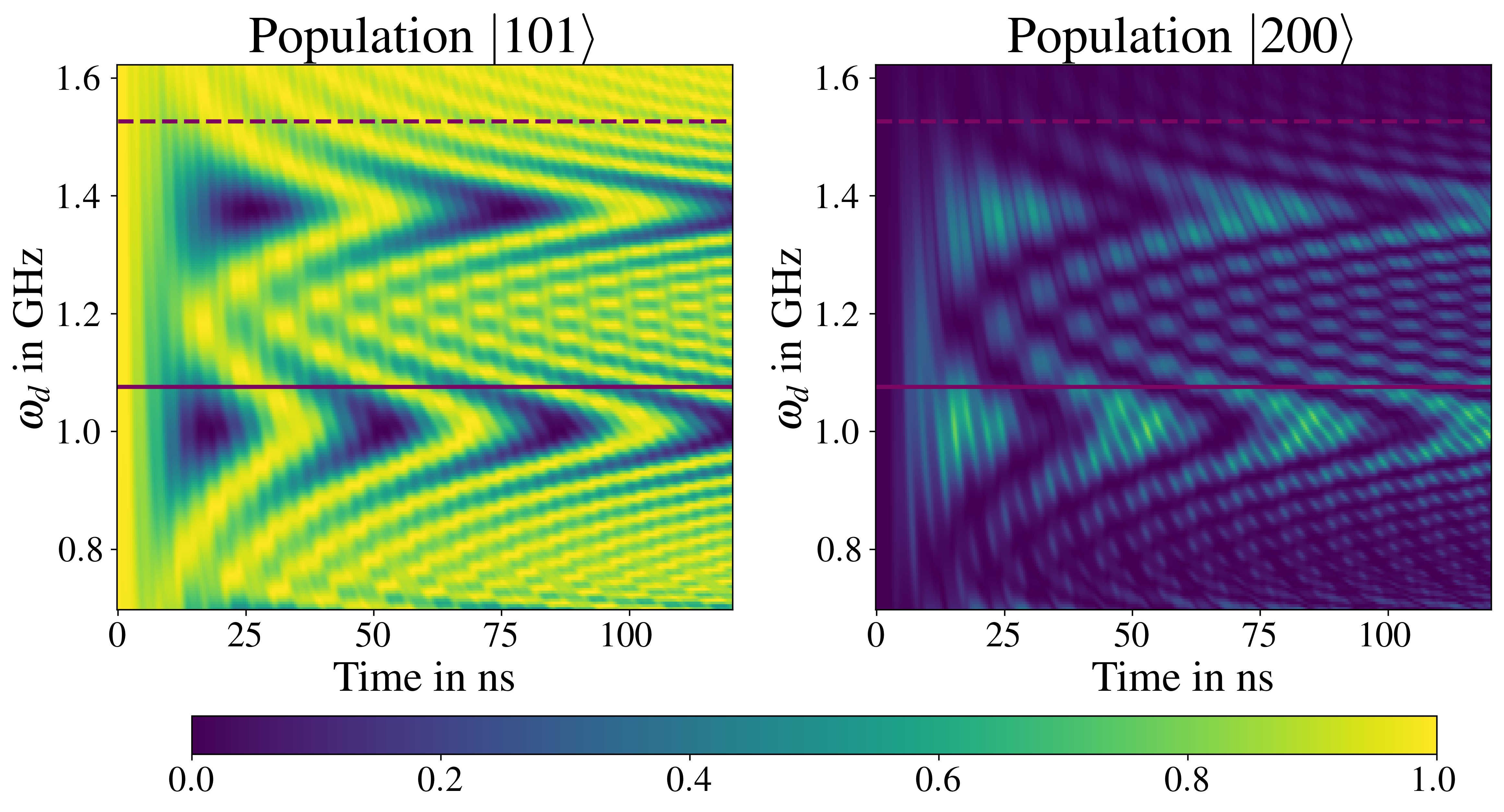}
    \caption{Chevron pattern for the transition of $|101\rangle$ to $|200\rangle$. The purple solid line denotes the optimized drive frequency $\omega_d$ for the 50 ns gate, while the dashed line corresponds to the correction frequency of the second drive tone.}
    \label{fig:Chevron}
\end{figure}

From this, we calculate the effective Hamiltonian in \autoref{eqn:Heff2ndord}. For a complete oscillation between $\ket{101}$ and $\ket{200}$, the corresponding diagonal entries of $\mathcal{H}_{\text{eff}}$ must be equal. This resonance condition determines the estimated drive frequency $\omega_D$, which, up to first order, is given by:
\begin{widetext}
\begin{align}
    \omega_D =& -\frac{\alpha}{4}\Biggl( \frac{E_{200}-E_{101}}{\alpha} + A_{22} - A_{00} - \notag \\ 
    &\sqrt{\biggl(\frac{E_{200}-E_{101}}{\alpha} + A_{22} - A_{00} \biggr)^2 + 8\biggl( \frac{9\beta^2+10\alpha^2}{30\alpha^2}A_{01}^2 + \frac{\beta^2+8\alpha^2}{8\alpha^2} A_{02}^2 + \frac{\beta^2-6\alpha^2}{6\alpha^2}A_{12}^2\biggr)}\Biggr) \ .
\end{align}
The exchange element of $\mathcal{H}_{\text{eff}}$ between states $\ket{101}$ and $\ket{200}$ denotes the effective coupling $g_{\text{eff}}$. This allows us to estimate the time for a full oscillation between both states:
\begin{align}
    t_{\text{gate}} = \frac{\pi}{|g_{\text{eff}}|} = \frac{\pi}{\frac{2\alpha \beta}{3\omega_D}A_{01}A_{12} + \left( \frac{1}{2} + \frac{\alpha(A_{22}-A_{00})}{4\omega_D}\right)\beta A_{02}} \ .
\end{align}
\end{widetext}
For short gate times, the drive frequency is detuned from resonance, as the additional $ZZ$ coupling arising from the avoided crossing of the $|110\rangle_F$ and $|200\rangle_F$ Floquet states significantly contributes to the selective phase accumulation on the $|101\rangle$ state. For that reason, the optimal drive tone for the CZ-gate shifts slightly away from the pure resonant transition as indicated in \autoref{fig:Chevron}.

\section{\label{sec:robustness}Robustness to fabrication imprecisions}
\begin{figure}[!htp]
    \centering
    \includegraphics[width=\columnwidth]{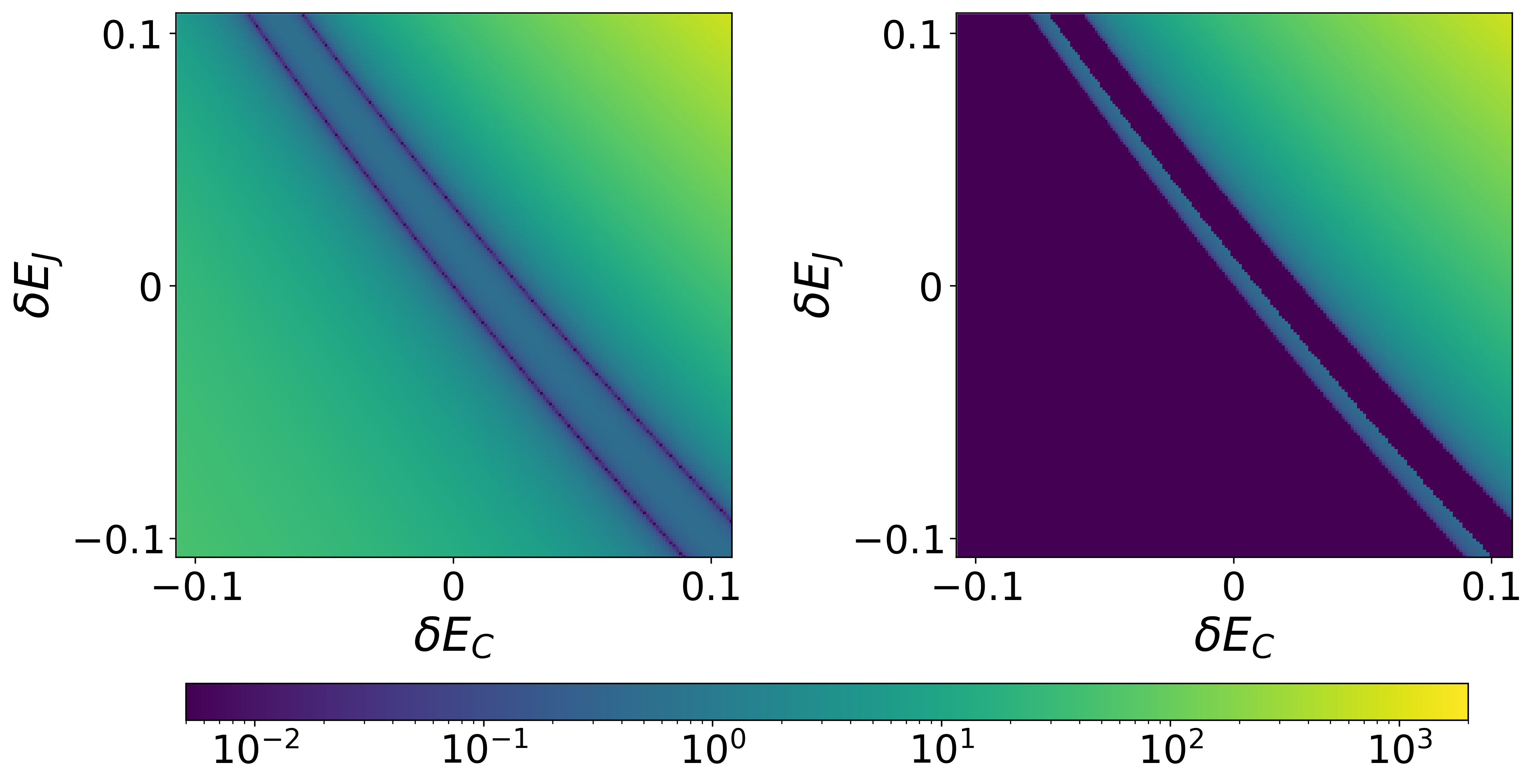}
    \caption{ZZ-crosstalk in kHz when considering errors in $E_C$ and $E_J$ in the coupler. We define the errors as $\delta E_X = (E_X - E_X^{\text{Target}})/E_X^{\text{Target}}$. The left plot shows the $ZZ$-crosstalk at the lower sweet spot of the coupler at $\phi_{\text{ext}, c} = \pi$, the right plot shows the minimal $ZZ$-coupling that can be achieved by readjusting the external flux $\phi_{\text{ext},c}$.}
    \label{fig:ZZcrosstalk}
\end{figure}

Achieving precise energy values for Josephson junctions remains a significant challenge in the fabrication of quantum circuits. The left plot in \autoref{fig:ZZcrosstalk}  illustrates that small deviations in the fabrication of the coupler’s $E_J$ and $E_C$ can result in minor $ZZ$-errors in the kHz regime. However, as shown in \autoref{fig:ZZcrosstalk}  on the right, these errors can often be mitigated by recalibrating the magnetic flux and adjusting the coupler slightly away from its lower sweet spot. This demonstrates the system’s robustness to fabrication inaccuracies, reinforcing its potential as a scalable architecture for quantum computing.

\begin{figure}[!htp]
    \centering
    \includegraphics[width=\columnwidth]{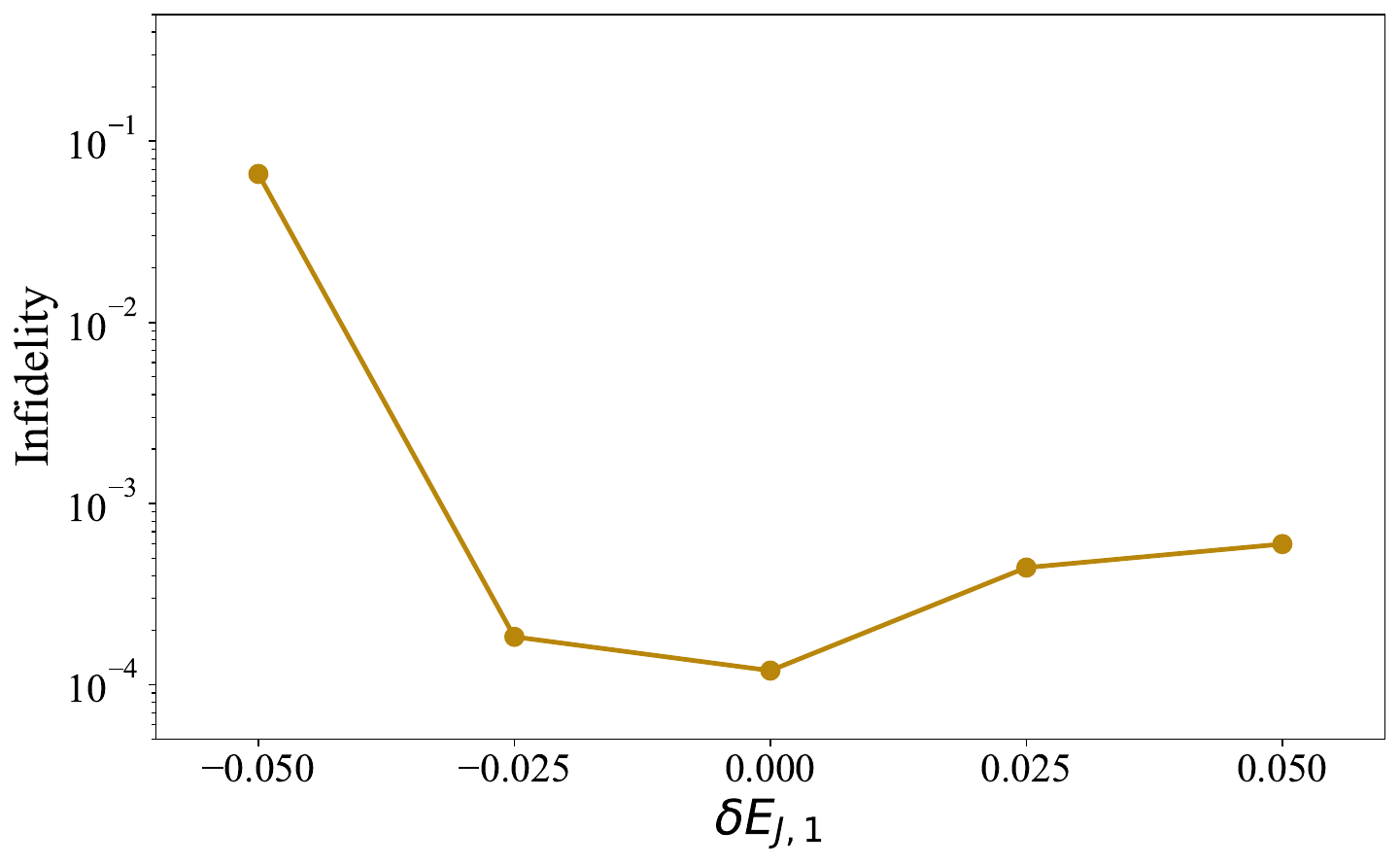}
    \caption{Simulated gate infidelities for a 40\,ns gate versus a fabrication error in the Josephson junction of the Fluxonium $E_{J,1}$}
    \label{fig:EJerr}
\end{figure}
To estimate the robustness of the parametrically driven CZ gate, we introduced a variation in the Josephson energy of the Fluxonium’s junction, as this circuit element is likely the most susceptible to fabrication imprecisions. \autoref{fig:EJerr} shows how the gate fidelity changes with a slightly smaller or larger junction. As seen in the plot, for small deviations, the gate infidelity remains on the same order of magnitude, indicating a degree of resilience to such imperfections.

\section{\label{sec:CZ_alternative}CZ-gate for tunable Transmons}
An alternative approach to implementing a CZ-gate in a hybrid Fluxonium-Transmon architecture involves making the ancilla Transmon flux-tunable using a dc SQUID. By adjusting the flux, the $\ket{101}$ state can be brought into resonance with the $\ket{200}$ state of the Fluxonium.

We simulated a $50\,\rm{ns}$ CZ gate using optimized Gaussian flattop pulse shapes on both the flux pulse of the coupler and the Transmon ancilla. The closed-system infidelity is $3.9 \times 10^{-5}$, while the estimated $T_1$ error is on the order of $10^{-3}$.

However, this approach presents a significant drawback: the ancilla Transmon must execute CZ-gates with all neighboring Fluxoniums during a stabilizer readout cycle. As a result, for certain CZ operations, the Transmon must be tuned across resonances with spectator qubits, increasing the risk of leakage errors due to interactions with these spectators. This issue is further exacerbated by the fact that the coupler is not biased to a point where it effectively suppresses the hopping rate between $\ket{101}$ and $\ket{200}$. Consequently, even when the coupler is nominally inactive, unintended population exchange within this subspace will still occur, leading to a high risk of executing unwanted gates.

\section{\label{sec:ZZLS} $ZZ$-crosstalk in a 3-qubit architecture}
\begin{figure}
    \centering
    \includegraphics[width=\columnwidth]{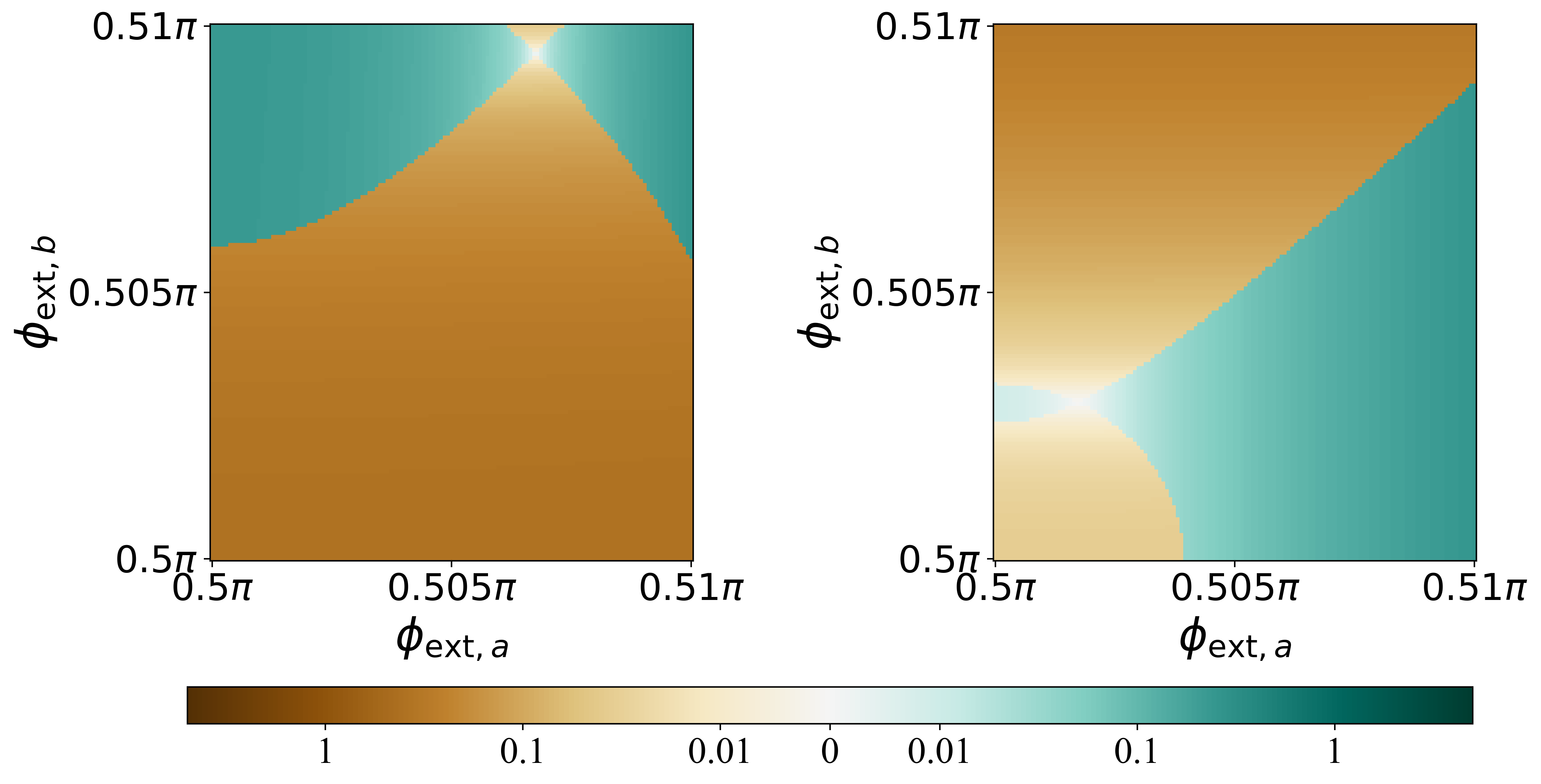}
    \caption{$ZZ$-crosstalk in kHz for the F-T-F configuration (left panel) and the T-F-T configuration (right panel). Brown regions indicate that the $ZZ$-crosstalk between qubits 1 and 2 dominates, while cyan regions indicate dominant $ZZ$-crosstalk between qubits 2 and 3. Although the plots show that the $ZZ$-crosstalk is generally in the (sub-)kHz regime, there exists a specific point in both configurations where the pairwise $ZZ$-crosstalk between both qubit pairs is simultaneously suppressed. This operating point can be reached by readjusting the magnetic fluxes through couplers $a$ and $b$.}
    \label{fig:ZZLS}
\end{figure}

\autoref{fig:ZZLS} shows the $ZZ$-crosstalk (in kHz) for both the F-T-F configuration (left) and the T-F-T configuration (right) as a function of the magnetic fluxes threading coupler $a$ and coupler $b$. The brown regions indicate that the $ZZ$-crosstalk between qubits 1 and 2 dominates, while the cyan regions indicate dominant crosstalk between qubits 2 and 3. As evident from the plots, tuning the magnetic fluxes within a small range keeps the $ZZ$-crosstalk within the (sub-)kHz regime, demonstrating the system’s resilience to flux noise. Moreover, each configuration features a specific operating point at which the pairwise $ZZ$-crosstalk is fully suppressed. Tuning the couplers to this point minimizes dephasing errors in the system.

